%% file: main_paper.tex
\newcommand*{\btag}{\ensuremath{{B}_{\text{tag}} }}
\newcommand*{\bsig}{\ensuremath{{B}_{\text{sig}} }}

\newcommand*{\R}{\ensuremath{\mathcal{R} }}
\newcommand*{\RD}{\ensuremath{\mathcal{R}(D) }}
\newcommand*{\RDSt}{\ensuremath{\mathcal{R}(D^{\ast}) }}
\newcommand*{\RDall}{\ensuremath{\mathcal{R}(D^{(\ast)}) }}

\newcommand*{\eecl}{\ensuremath{E_{\text{ECL}} }}
\newcommand*{\costheta}{\ensuremath{\cos\theta_{B,D^{(\ast)}\ell} }}

\newcommand*{\tauDec}{\ensuremath{\tau^- \to \ell^- \bar{\nu}_{\ell} \nu_{\tau} }}

\documentclass[aps,prd,twocolumn,superscriptaddress,showpacs,preprintnumbers,amsmath,amssymb]{revtex4-1}

\usepackage{graphicx}
\usepackage{dcolumn}
\usepackage{bm}
\usepackage{hyperref}
\usepackage[mathlines]{lineno}
\usepackage{subfigure}
\usepackage{multirow}
\usepackage{color}
\usepackage{url}
\usepackage{xcolor}

\hypersetup{
}

\graphicspath{{imports/figures/}}

\begin{document}

\preprint{\vbox{ 
                  \hbox{Belle-CONF-1902}
}}

\title{Measurement of \RD\ and \RDSt\ with a semileptonic tagging method}



\date{\today}

\input{imports/authors.tex}

\begin{abstract}
We report a measurement of the ratios of branching fractions $\RD = {\cal B}(\bar{B} \to D \tau^- \bar{\nu}_{\tau})/{\cal B}(\bar{B} \to D \ell^- \bar{\nu}_{\ell})$ and $\RDSt = {\cal B}(\bar{B} \to D^* \tau^- \bar{\nu}_{\tau})/{\cal B}(\bar{B} \to D^* \ell^- \bar{\nu}_{\ell})$, where $\ell$ denotes an electron or a muon. The results are based on a data sample containing $772\times10^6$ $B\bar{B}$ events recorded at the $\Upsilon(4S)$ resonance with the Belle detector at the KEKB $e^+ e^-$ collider. The analysis utilizes a method where the tag-side $B$ meson is reconstructed in a semileptonic decay mode, and the signal-side $\tau$ is reconstructed in a purely leptonic decay. The measured values are $\RD\ = 0.307 \pm 0.037 \pm 0.016$ and $\RDSt\ = 0.283 \pm 0.018 \pm 0.014$, where the first uncertainties are statistical and the second are systematic. These results are in agreement with the Standard Model predictions within $0.2$ and $1.1$ standard deviations, respectively, while their combination agrees with the Standard Model predictions within $1.2$ standard deviations.  
\end{abstract}


\maketitle


\section{INTRODUCTION}
Semitauonic $B$ meson decays of the type $b \to c \tau \nu_{\tau}$~\cite{note} are sensitive probes for physics beyond the Standard Model (SM). Charged Higgs bosons, which appear in supersymmetry~\cite{Martin:1997ns} and other models with two Higgs doublets~\cite{Gunion:1989we}, may contribute measurably to the decays due to the large masses of the $\tau$ and the $b$ quark. Similarly, leptoquarks~\cite{Buchmuller:1986zs}, which carry both baryon and lepton numbers, may also contribute to this process. The ratio of branching fractions
\begin{eqnarray}
    \RDall = \frac{{\cal B}(\bar{B} \to D^{(*)} \tau^- \bar{\nu}_{\tau})}{{\cal B}(\bar{B} \to D^{(*)} \ell^- \bar{\nu}_{\ell})} \hspace{0.8em}(\ell = e,\mu)
\end{eqnarray}
is typically measured instead of the absolute branching fraction of $\bar{B} \to D^{(*)} \tau^- \bar{\nu}_{\tau}$ to reduce common systematic uncertainties, such as those on the detection efficiency, the magnitude of the quark-mixing matrix element $|V_{cb}|$, and the semileptonic decay form factors. Hereafter, $\bar{B} \to D^{(*)} \tau^- \bar{\nu}_{\tau}$ and $\bar{B} \to D^{(*)} \ell^- \bar{\nu}_{\ell}$ will be referred to as the signal and normalization modes, respectively. The SM calculations for these ratios, performed by several groups~\cite{Bigi:2016mdz, Bernlochner:2017jka, Bigi:2017jbd, Jaiswal:2017rve}, are averaged \cite{HFLAV} to obtain
$\RD  = 0.299 \pm 0.003$ and
$\RDSt    = 0.258 \pm 0.005$.

Semitauonic $B$ decays were first observed by Belle~\cite{PhysRevLett.99.191807}, with subsequent studies reported by Belle~\cite{,PhysRevD.82.072005, huschle2015, Sato:2016svk, Hirose:2016wfn}, BaBar~\cite{lees2012}, and LHCb~\cite{aaij2015, Aaij:2017deq}.
The average values of the experimental results are
$\RD = 0.407 \pm 0.039 \pm 0.024$ and
$\RDSt = 0.306 \pm 0.013 \pm 0.007$~\cite{HFLAV},
where the first uncertainty is statistical and the second is systematic. These values exceed SM predictions by $2.1\sigma$ and $3.0\sigma$, respectively. Here, $\sigma$ denotes the standard deviation. A combined analysis of \RD\ and \RDSt\, taking correlations into account, finds that the deviation from the SM prediction is approximately $3.8\sigma$~\cite{HFLAV}.

So far,  simultaneous measurements of \RD\ and \RDSt\ at the ``B factory" experiments Belle and BaBar have been performed using hadronic tagging methods and both $B^0$ and $B^+$ decays~\cite{huschle2015,lees2012}, while only $\R(D^{*+})$ was measured with a semileptonic tag~\cite{Sato:2016svk}. In this paper, we report the first measurement of \RD\ using the semileptonic tagging method, and update our earlier measurement of \RDSt\ by combining results of $B^0$ and $B^+$ decays using a more efficient tag reconstruction algorithm.


\section{DETECTOR AND MC SIMULATION}
We use the full $\Upsilon(4S)$ data sample containing $772 \times 10^6$ $B \bar{B}$ events recorded with the Belle detector~\cite{Abashian:2000cg} at the KEKB $e^+ e^-$ collider~\cite{KUROKAWA20031}.
Belle is a general-purpose magnetic spectrometer, which consists of a silicon vertex detector (SVD), a 50-layer central drift chamber (CDC), an array of aerogel threshold Cherenkov counters (ACC), time-of-flight scintillation counters (TOF), and an electromagnetic calorimeter (ECL) comprising CsI(Tl) crystals. These components are located inside a superconducting solenoid coil that provides a 1.5 T magnetic field. An iron flux-return yoke located outside the coil is instrumented to detect $K_L^0$ mesons and muons (KLM). The detector is described in detail elsewhere~\cite{Abashian:2000cg}.
 
To determine the reconstruction efficiency and probability density functions (PDFs)  for signal, normalization, and background modes, we use Monte Carlo (MC) simulated events, generated with the EvtGen event generator~\cite{Ryd:2005zz}, and having the detector response simulated with the GEANT3 package~\cite{Brun:1987ma}. 

The $B \to D^{(*)} \ell \nu$ decays are generated with the \verb+HQET2+ EvtGen package, based on the CLN parametrization~\cite{Caprini1997}. As parameters of the model have been updated since our MC generation, we apply an event-by-event correction factor obtained by taking the ratio of the momentum transfer $q^2$ and lepton momentum $p^*_\ell$ in the centre-of-mass frame of the B meson distributions of the new and old model. On the other hand, MC samples for the $B \to D^{**} \ell \nu$ decays are generated with the \verb+ISGW2+ EvtGen package, based on the quark model described in Ref.~\cite{Scora1995}. This model has been superseded by the LLSW model~\cite{Leibovich1998}; thus we weight events with a correction factor based on the ratio of the analytic predictions of LLSW and MC distributions generated with \verb+ISGW2+. Here, $D^{**}$ denotes the orbitally excited states $D_1$, $D_2^*$, $D_1'$, and $D_0^*$. 

We consider $D^{**}$ decays to a $D^{(*)}$ and a pion, a $\rho$ or an $\eta$ meson, or a pair of pions, where branching fractions are based on quantum number, phase-space, and isospin arguments. The sample sizes of the generic $\Upsilon(4S) \to B\bar{B}$ and $B \to D^{**} \ell \nu $ processes correspond to about 10 times and 5 times the integrated luminosity of the $\Upsilon(4S)$ data sample, respectively.


\section{EVENT RECONSTRUCTION AND SELECTION}
The \btag\ is reconstructed using a hierarchical algorithm based on ``Fast" boosted decision trees (BDT)~\cite{Keck2019} in the $D^{} \ell \bar{\nu}_{\ell}$ and $D^{*} \ell \bar{\nu}_{\ell}$ channels, where $\ell = e, \mu$. We select well-reconstructed \btag\ candidates by requiring their classifier output to be larger than $10^{-1.5}$. We veto ${B} \to D^{\ast} \tau (\to \ell \nu \nu) \nu  $ events on the tag side by applying a selection on  \costheta\ . This variable represents the cosine of the angle between the momentum of the $B$ meson and the $D^{(*)} \ell$ system in the $\Upsilon(4S)$ rest frame, under the assumption that only one massless particle is not reconstructed:
\begin{eqnarray}
     \costheta \equiv
    \frac
    {2E_{\rm beam} E_{D^{(*)} \ell} - m_B^2 - m_{D^{(*)} \ell}^2}
    {2 |\bm{p}_B| |\bm{p}_{D^{(*)} \ell}|},
    \label{eq:cos_bdstrl}
\end{eqnarray}
where $E_{\rm beam}$ is the beam energy, and $E_{D^* \ell}$, $\bm{p}_{D^* \ell}$, and $m_{D^* \ell}$ are the energy, momentum, and mass, respectively, of the $D^* \ell$ system. The quantity $m_B$ is the nominal $B$ meson mass~\cite{PDG}, and $\bm{p}_B$ is the $B$ meson momentum. All quantities are evaluated in the $\Upsilon(4S)$ rest frame.  

Correctly reconstructed $B$ candidates in the normalization mode are expected to have a value of  \costheta\ between $-1$ and $+1$. Similarly, correctly reconstructed and misreconstructed $B$ candidates in the signal mode tend to have \costheta\ values more negative than this range due to additional missing particles. We account for detector resolution effects and apply the requirement $ -2.0 <  \costheta < 1.0$ for the \btag.

In each event with a selected \btag\ candidate, we search for the signature $D^{(*)} \ell$ among the remaining tracks and calorimeter clusters. We define four disjoint data samples, denoted $D^{+} \ell^{-}$, $D^{0} \ell^{-}$, $D^{*+} \ell^{-}$, and $D^{*0} \ell^{-}$.

Charged particle tracks are reconstructed with the SVD and CDC by requiring a point of closest approach to the interaction point smaller than 5.0 (2.0) cm along (transverse to) the $z$ axis. Here, the $z$ axis is opposite the $e^+$ beam direction. These requirement do not apply to pions daughters from $K_S^0$ decays. Electrons are identified by a combination of the specific ionization ($dE/dx$) in the CDC, the ratio of the cluster energy in the ECL to the track momentum measured with the CDC, the response of the ACC, the cluster shape in the ECL, and the match between positions of the cluster and the track at the ECL. To recover bremsstrahlung photons from electrons, we add the four-momentum of each photon detected within 0.05 rad of the original track direction to the electron momentum. Muons are identified by the track penetration depth and hit distribution in the KLM. Charged kaons are identified by combining information from the $dE/dx$ measured in the CDC,  the flight time measured with the TOF, and the response of the ACC. We do not apply any particle identification criteria for charged pions.

Candidate $K_S^0$ mesons are formed by combining two oppositely charged tracks with pion mass hypotheses. We require their invariant mass to lie within $\pm$15 MeV/$c^2$ of the nominal $K^0$ mass~\cite{PDG}, which corresponds to approximately 7 times the reconstructed mass resolution. Further selection is performed with an algorithm based on a NeuroBayes neural network~\cite{FEINDT2006190}.

Photons are measured as an electromagnetic cluster in the ECL with no associated charged track. Neutral pions are reconstructed in the $\pi^0 \to \gamma \gamma$ channel, and their energy resolution is improved by performing a mass-constrained fit of the two photon candidates to the nominal  $\pi^0$  mass~\cite{PDG}. For neutral pions from $D$ decays, we require the daughter photon energies to be greater than 50 MeV, the cosine of the angle between two photons to be greater than zero, and the $\gamma \gamma$ invariant mass to be within $[-15, +10]$ MeV/$c^2$ of the nominal $\pi^0$ mass, which corresponds to approximately $\pm 1.8$ times the resolution.  Low energy  $\pi^0$  candidates from $D^*$ are reconstructed using looser energy requirements: one photon must have an energy of at least 50 MeV, while the other must have a minimum energy of 20 MeV.  We also require a narrower window around the diphoton invariant mass to compensate for the lower photon-energy requirement: within 10 MeV/$c^2$ of the nominal $\pi^0$ mass, which corresponds to approximately $\pm 1.6$  times the resolution.

Neutral $D$ mesons are reconstructed in the following decay modes:
$D^0 \to K^- \pi^+ \pi^0$,
$K^- \pi^+ \pi^+ \pi^- $,
$K^- \pi^+$,
$K_S^0 \pi^+ \pi^-$,
$K_S^0 \pi^0$,
$K_S^0 K^+ K^-$,
$K^- K^+$ and
$\pi^- \pi^+$.
Similarly, charged $D$ mesons are reconstructed in the following modes:
$D^+ \to K^- \pi^+ \pi^+$,
$K_S^0 \pi^+ \pi^0$,
$K_S^0 \pi^+ \pi^+ \pi^-$,
$K_S^0 \pi^+$,
$K^- K^+ \pi^+$ and
$K_S^0 K^+$.
The combined branching fractions for reconstructed channels are 30\% and 22\% for $D^0$ and $D^+$, respectively. For $D$ decays without a $\pi^0$ in the final state,  we require the invariant mass of the reconstructed candidates to be within 15 MeV/$c^2$ of the $D^0$ or $D^+$ mass, which corresponds to a window of approximately $\pm 3$ times the resolution. In case of channels with a $\pi^0$ in the final state, which exhibit a worse mass resolution, we require a wider window: from $-45$ to $+30$ MeV/$c^2$ around the nominal $D^0$ mass, and  from $-36$ to $+24$ MeV/$c^2$ around the nominal $D^+$ mass. These windows correspond to approximately [$-1.2, +1.8$] and [$-1.0, +1.5$] times the resolution, respectively. Candidate $D^{*+}$ mesons are reconstructed in the channels $D^0 \pi^+$  and $D^+ \pi^0$, and $D^{*0}$ for the channel $D^0 \pi^0$. We do not consider the $D^{*0} \to D^0 \gamma$ decay channel due to higher backgrounds.

To improve the resolution of the $D^*$-$D$ mass difference, $\Delta M$, for the $D^{*+} \to D^0 \pi^+$ decay mode, the charged pion track from the $D^{*+}$ is refitted to the $D^0$ decay vertex. We require $\Delta M$ be within 2.5 MeV/$c^2$  around the nominal $D^*$-$D$ mass difference for the $D^{*+} \to D^0 \pi^+$ decay mode, and within 2.0 MeV/$c^2$ for the $D^{*+} \to D^+ \pi^0$ and $D^{*0} \to D^0 \pi^0$ decay modes. These windows correspond to $\pm 3.2$ and $\pm 2.0$ times the resolution, respectively. We require a tighter mass window in the $D^{*}$ modes containing low-momentum (``slow") $\pi^0$ to suppress a large background arising from misreconstructed neutral pions.

In each event we require that there be two $B$ candidates of opposite in flavor.  While it is possible for signal events to have the same flavor due to $B \bar{B}$ mixing, we do not allow such events as they lead to ambiguous $D^* \ell$ pair assignment and hence to a larger combinatorial background. 

On the signal side, we require \costheta\ to be less than 1.0 and the $D^{(*)}$ momentum in the $\Upsilon(4S)$ rest frame to be less than 2.0 GeV/$c$.
Finally, we require that events contain no extra charged tracks, $K_S^0$ candidates, or $\pi^0$ candidates, which are reconstructed with the same criteria as those used for the $D$ candidates.

When multiple  \btag\ or  \bsig\ candidates are found in an event, we select the \btag\ candidate with the highest tagging classifier output, and the \bsig\ candidate with the highest p-value resulting from the $D$ or $D^*$ vertex fit. The efficiencies of the best candidate selection algorithm are 95\%, 93\%, 88\%, and 86\% for the $D^{+} \ell^{-}$, $D^{0} \ell^{-}$, $D^{*+} \ell^{-}$ and $D^{*0} \ell^{-}$ samples, respectively.
%
\section{SIGNAL EXTRACTION}

\begin{figure}[htb]
  \includegraphics[width=\columnwidth]{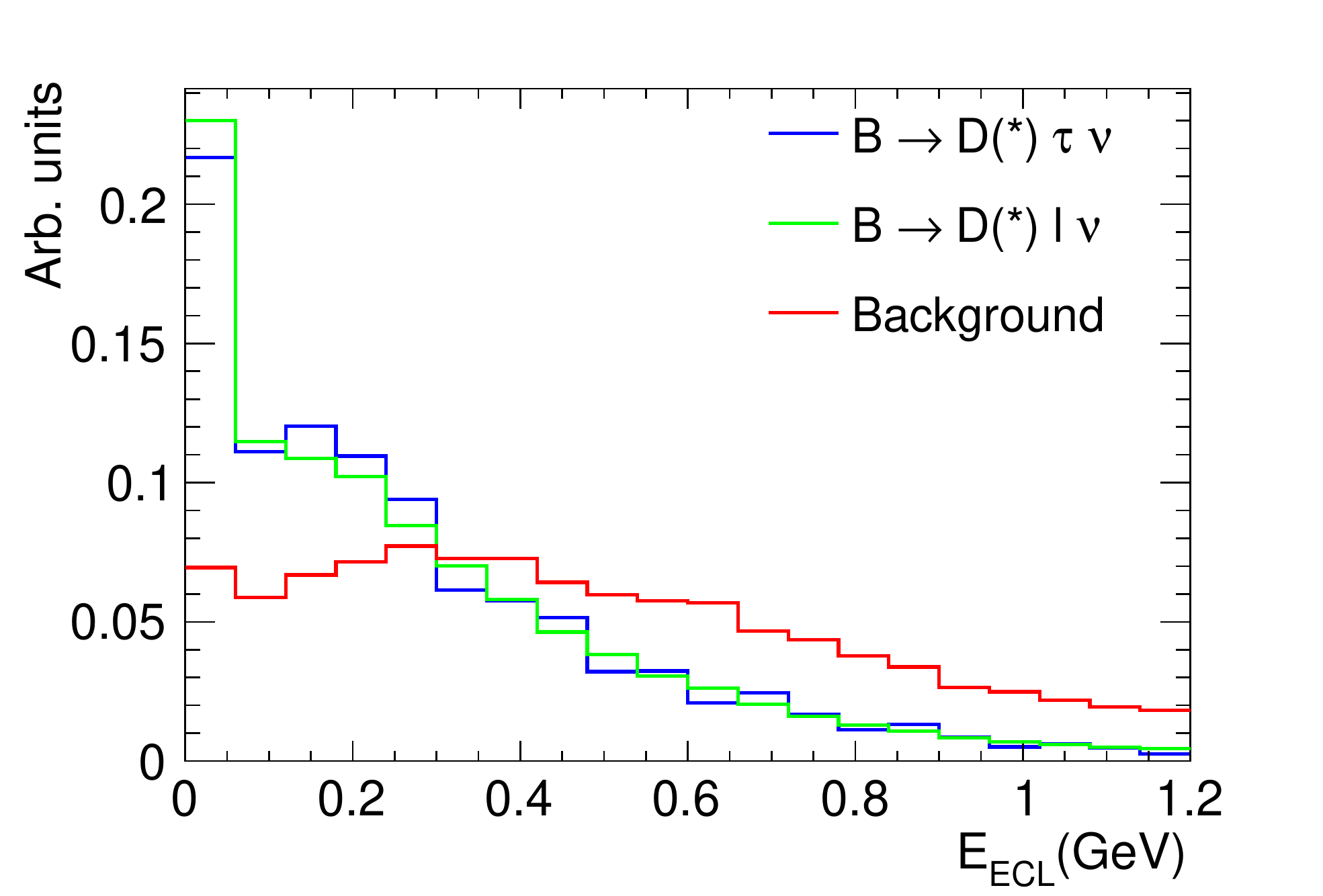}
  \caption{\eecl\ distributions for the signal, normalization, and background taken from MC simulation. The distributions for all decay modes are summed together and normalized to unity.}
  \label{fig:th_MC_eecl}
\end{figure}

To distinguish signal and normalization events from background processes, we use the sum of the energies of neutral clusters detected in the ECL that are not associated with reconstructed particles, denoted as \eecl.
To mitigate the effects of photons related to beam background, for the \eecl\ calculation we include only clusters with energies greater than 50, 100, and 150 MeV, respectively, from the barrel, forward, and backward calorimeter regions~\cite{Abashian:2000cg}.
Signal and normalization events peak near zero in $E_{\rm ECL}$, while background events populate a wider range as shown in Figure~\ref{fig:th_MC_eecl}. We require that \eecl\ be less than 1.2 GeV.

To separate reconstructed signal and normalization events, we employ a BDT based on the \verb|XGBoost| package~\cite{xgboost16}. The input variables to the BDT are  \costheta; the approximate missing mass squared $m_{\rm miss}^2 = ( E_{\text{beam}} - E_{D^{(*)}} - E_\ell )^2 - (\bm{p}_{D^{(*)}} + \bm{p}_{\ell} )^2$; the visible energy $E_{\rm vis} = \sum_i E_i$, where $(E_i, \bm{p}_i)$ is the four-momentum of particle $i$. The BDT classifier is trained for each of the four $D^{(*)} \ell$ samples using MC events of signal and normalization modes. We do not apply any selection on the BDT classifier output, denoted as \verb|class|; instead we use it as one of the fitting variables for the extraction of \RDall.

%


We extract the yields of signal and normalization modes from a two-dimensional (2D) extended maximum-likelihood fit to the variables \verb|class| and \eecl. The fit is performed simultaneously to the four $D^{(*)} \ell$ samples. The distribution of each sample is described as the sum of several components: $D^{(*)} \tau \nu$,  $D^{(*)} \ell \nu$, feed-down from $D^* \ell (\tau) \nu$ to $D \ell (\tau) \nu$, $D^{**} \ell/\tau \nu$, and other backgrounds.  The PDFs of these components are determined from MC simulations.
A large fraction of $B \to D^* \ell \nu$ decays for  both $B^0$ and $B^+$ is reconstructed in the $D \ell$ samples (feed-down). We leave these two contributions free in the fit and use their fitted yields to estimate the feed-down rate of $B \to D^* \tau \nu$ decays. As the probability of $B \to D (\ell/\tau) \nu$ decays contributing to the $D^* \ell$ samples is small, the rate of this contribution is fixed to its expected value.

The free parameters in the final fit are the yields of signal, normalization, $B \to D^{**} \ell \nu_{\ell}$ and feed-down from $D^* \ell$ to $D \ell$ components. The yield of fake $D^{(*)}$ events is fixed to the value estimated from the $\Delta M$ sidebands. The yields of other backgrounds are fixed to their MC expected values.
The ratios \RDall\ are given by the formula:
\begin{eqnarray}
    \RDall &=&
    \frac{1}{2{\cal B}(\tau^- \to \ell^- \bar{\nu}_{\ell} \nu_{\tau})}
    \cdot
    \frac{\varepsilon_{\rm norm}}{\varepsilon_{\rm sig}}
    \cdot
    \frac{N_{\rm sig}}{N_{\rm norm}},
    \label{eq:cal_rdstr}
\end{eqnarray}
where $\varepsilon_{\rm sig (norm)}$ and $N_{\rm sig (norm)}$ are the detection efficiency and yields of signal (normalization) modes and ${\cal B}(\tau^- \to \ell^- \bar{\nu}_{\ell} \nu_{\tau})$ is the average of the world averages for $\ell =e$ and $\ell = \mu$.


To improve the accuracy of the MC simulation, we apply a series of correction factors determined from control sample measurements. The lepton identification efficiencies are separately corrected for electrons and muons to account for differences between data and simulations  in the detector responses. Correction factors for these efficiencies are evaluated as a functions of the lepton momentum and direction using $e^+ e^- \to e^+ e^- \ell^+ \ell^-$ and $J/\psi \to \ell^+ \ell^-$ decays. 

We reweight events to account for differing yields of misreconstructed $D^{(*)}$  between data and MC simulations. The calibration factor for the fake charm correction is provided by the ratio of 2D histograms of  \verb|class| vs. \eecl\ for the $\Delta M$ sideband of data and MC events. 
In order to correct for the difference in \btag\ reconstruction efficiencies between data and MC simulations, we build PDFs of correctly reconstructed and misreconstructed \btag\ candidates using MC samples, and perform a fit to data. The ratios between the measured and expected yields provide the \btag\ calibration factors.
To validate the fit procedure, we perform fits to multiple subsets of the available MC samples. We do not find any bias with the evaluation of the statistical uncertainties.


\section{SYSTEMATIC UNCERTAINTIES}

To estimate various systematic uncertainties contributing to \RDall, we vary each fixed parameter 500 times, sampling from a Gaussian distribution built using the parameter's value and uncertainty. Then we repeat the fit and estimate the associated systematic uncertainty from the standard deviation of the resulting distribution. The systematic uncertainties are summarized in Table~\ref{tab:sys}. 

In Table~\ref{tab:sys} the label ``$D^{**}$ composition" refers to the uncertainty introduced by the branching fractions of the $B \to D^{**} \ell \nu_{\ell}$ channels and the decays of the $D^{**}$ mesons, which are not well known and hence contribute significantly to the total PDF uncertainty due to $B \to D^{**} \ell \nu_{\ell}$ decays. The uncertainties on the branching fraction of $B \to D^{**} \ell \nu_{\ell}$ are assumed to be $\pm 6\%$ for $D_1$, $\pm 10\%$ for $D_2^*$, $\pm 83\%$ for $D_1'$, and $\pm 100\%$ for $D_0^*$, while the uncertainties on each of the $D^{**}$ decay branching fractions are conservatively assumed to be $\pm 100\%$.

The efficiency factors for the fake $D^{(*)}$ and \btag\ reconstruction are calibrated using collision data. The uncertainties on these factors is affected by the size of the samples used in the calibration. We vary the factors within their errors and extract associated systematic uncertainties.


The reconstruction efficiency of feed-down events, together with the efficiency ratio of signal to normalization events, are varied within their uncertainties, which are limited by the size of MC samples.

The effect of the lepton efficiency and fake rate, as well as that due to the slow pion efficiency, do not cancel out in the $\R (D^{(*)})$ ratios. This is due to the different momentum spectra of leptons and charm mesons in the normalization and signal modes. The uncertainties introduced by these factors are included in the total systematic uncertainty.

A large systematic uncertainty arises from the limited size of MC samples. To estimate it, we recalculate PDFs for signal, normalization, fake $D^{(*)}$ events, $B \to D^{**} \ell \nu_{\ell}$, feed-down, and other backgrounds by generating toy MC samples from the nominal PDFs according to a Poisson statistics, and then repeat the fit with the new PDFs.

We include minor systematic contributions from other sources: one related to the parameters that are used for reweighting the semileptonic $B$ decays from the ISGW to LLSW model; and the others from the integrated luminosity and the branching fractions of $B \to D^{(*)} \ell \nu, D, D^*$ and  $\tau^- \to \ell^- \bar{\nu}_{\ell} \nu_{\tau}$ decays~\cite{PDG}.
The total systematic uncertainty is estimated by summing the aforementioned contributions in quadrature.

\begin{table}
	\caption{Systematic uncertainties contributing to the \RDall results.}
	\begin{ruledtabular}
	\centering
	\begin{tabular}{lcc}
		Source 								    &  $\Delta R(D) \ (\%)$    &  $\Delta R(D^{*}) \ (\%)$ \\
		\hline
		$D^{**}$ composition                    &                     0.76 &                      1.41 \\
		Fake $D^{(*)}$ calibration              &                     0.19 &                      0.11 \\
		\btag\ calibration                      &                     0.07 &                      0.05 \\
		Feed-down factors                       &                     1.69 &                      0.44 \\
		Efficiency factors                      &                     1.93 &                      4.12 \\
		Lepton efficiency and fake rate         &                     0.36 &                      0.33 \\
		Slow pion efficiency                    &                     0.08 &                      0.08 \\
		MC statistics                           &                     4.39 &                      2.25 \\
		$B$ decay form factors                  &                     0.55 &                      0.28 \\
		Luminosity                              &                     0.10 &                      0.04 \\
		$\mathcal{B}(B \to D^{(*)} \ell \nu)$   &                     0.05 &                      0.02 \\
		$\mathcal{B}(D)$                        &                     0.35 &                      0.13 \\
		$\mathcal{B}(D^*)$                      &                     0.04 &                      0.02 \\
		$\mathcal{B}(\tauDec)$ 				    &                     0.15 &                      0.14 \\
		\hline
		Total                                	&                     5.21 &                      4.94 \\
	\end{tabular}
	\label{tab:sys}
	\end{ruledtabular}
\end{table}


\section{RESULTS}

\begin{figure*}[tbp]
	\includegraphics[width=\columnwidth]{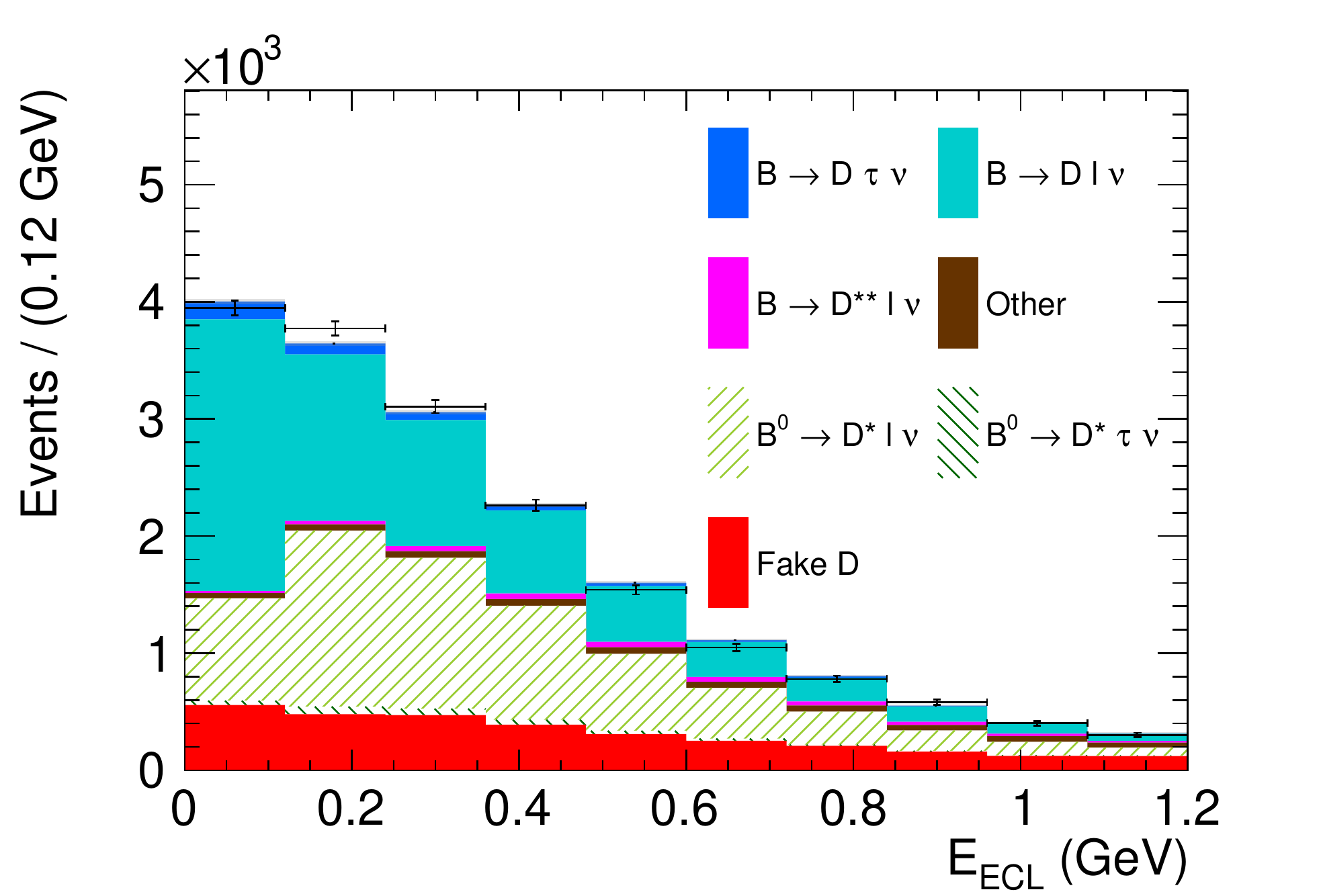}
	\includegraphics[width=\columnwidth]{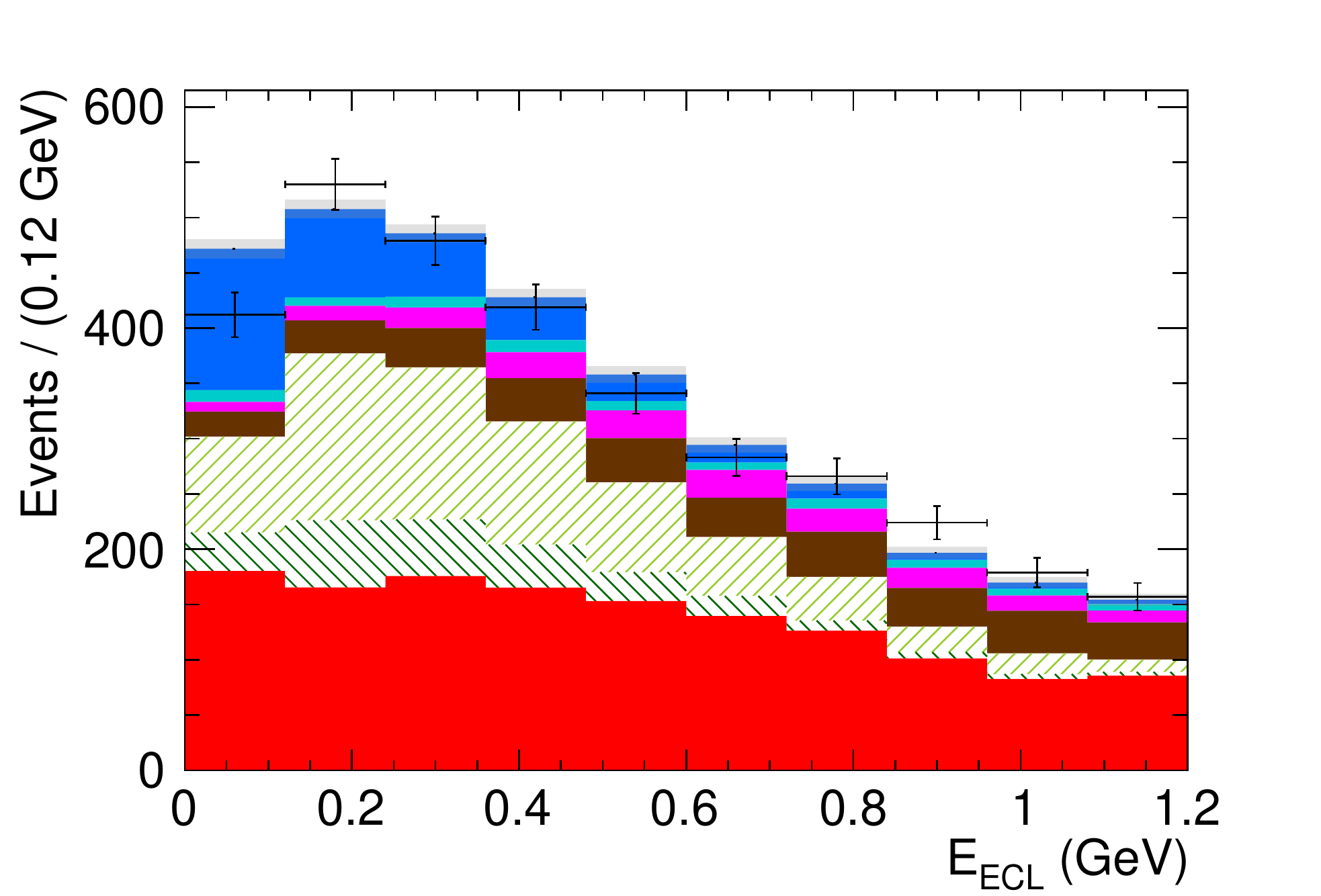}
	\includegraphics[width=\columnwidth]{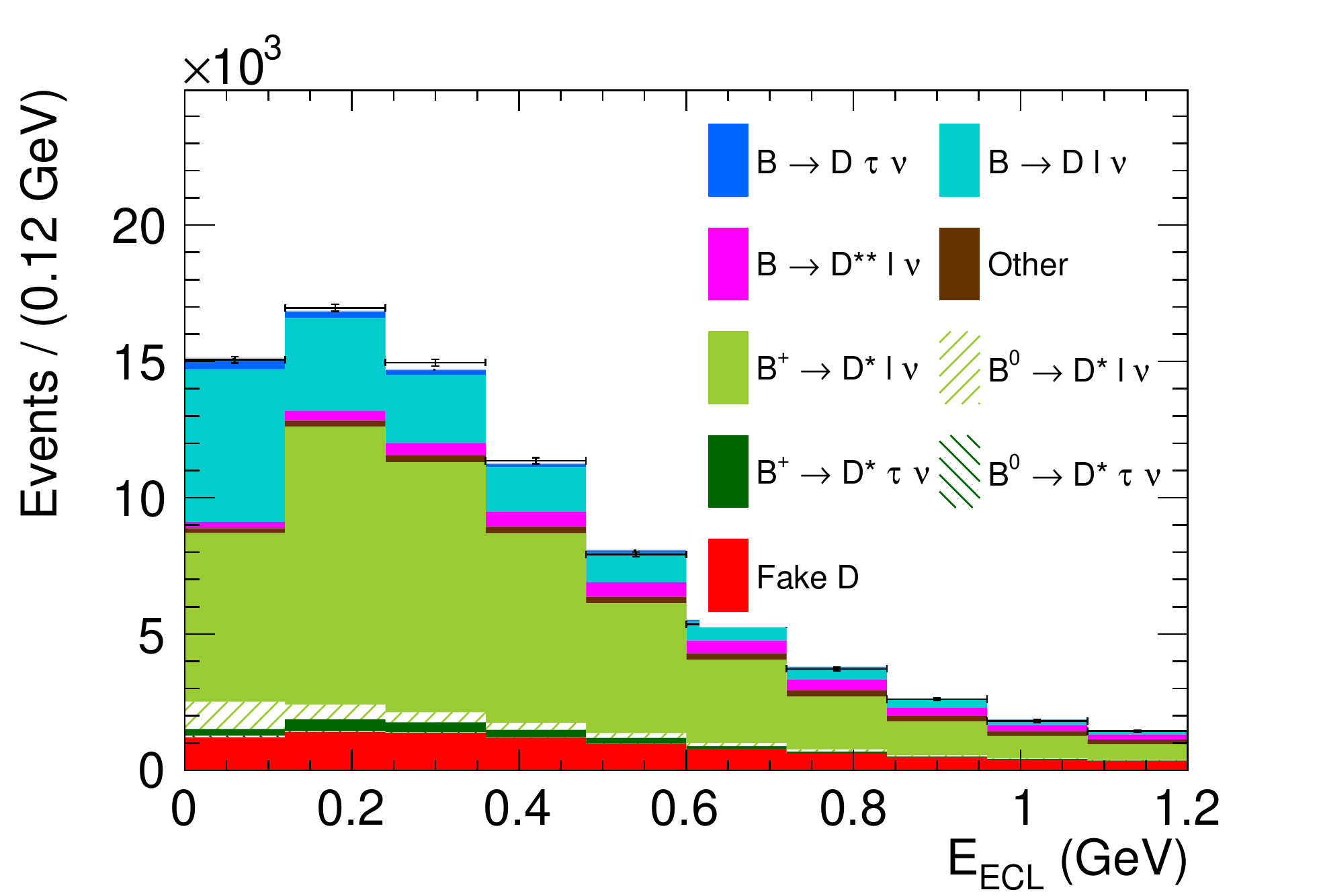}
	\includegraphics[width=\columnwidth]{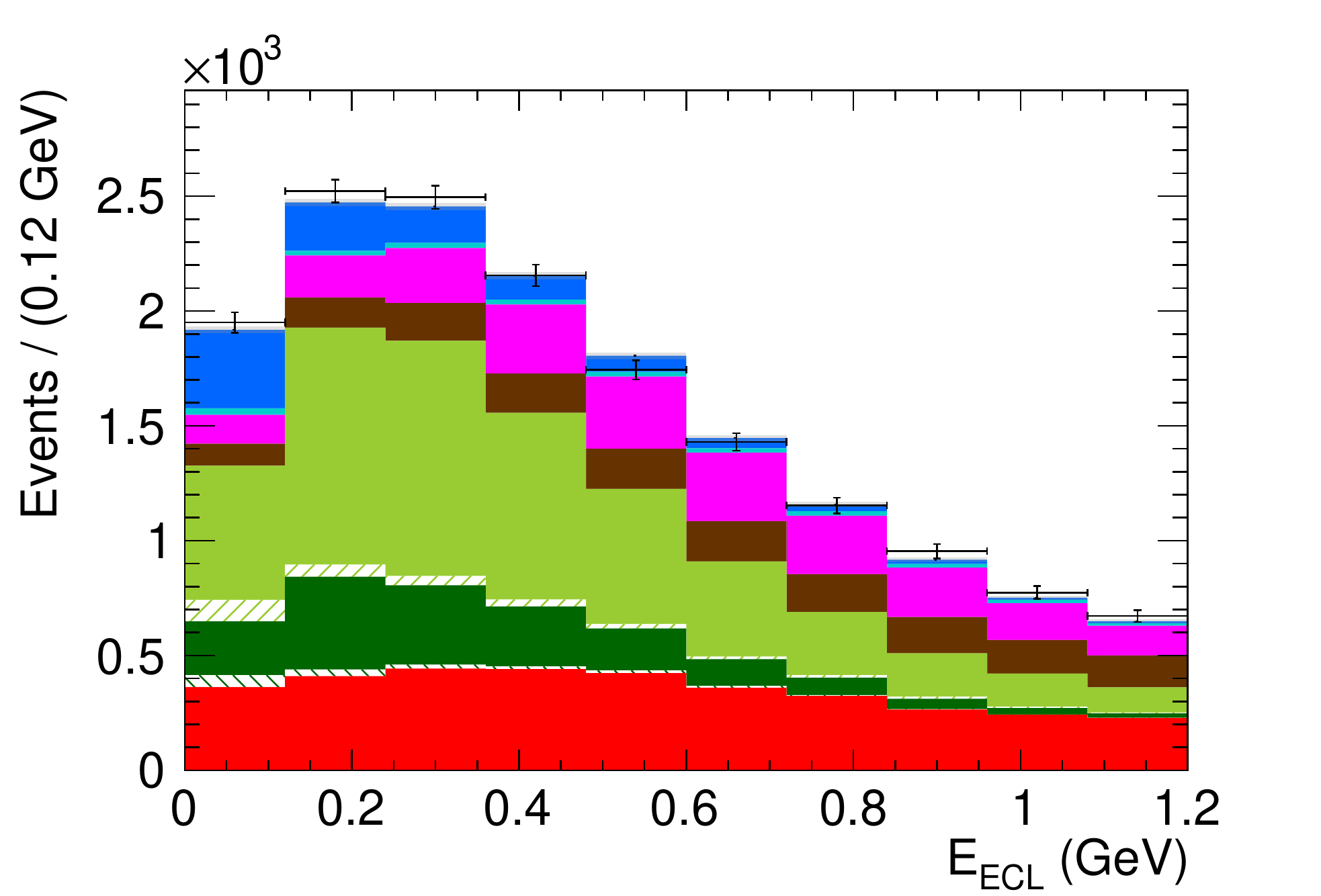}
\caption{\eecl\ fit projections and data points with statistical uncertainties in the $D^+\ell^-$ (top) and $D^0\ell^-$ (bottom) samples, for the full classifier region (left) and the signal region defined by the selection \texttt{class} $>0.9$ (right).}
\label{fig:results_Dmodes}
\end{figure*}

\begin{figure*}[tbp]
	\includegraphics[width=\columnwidth]{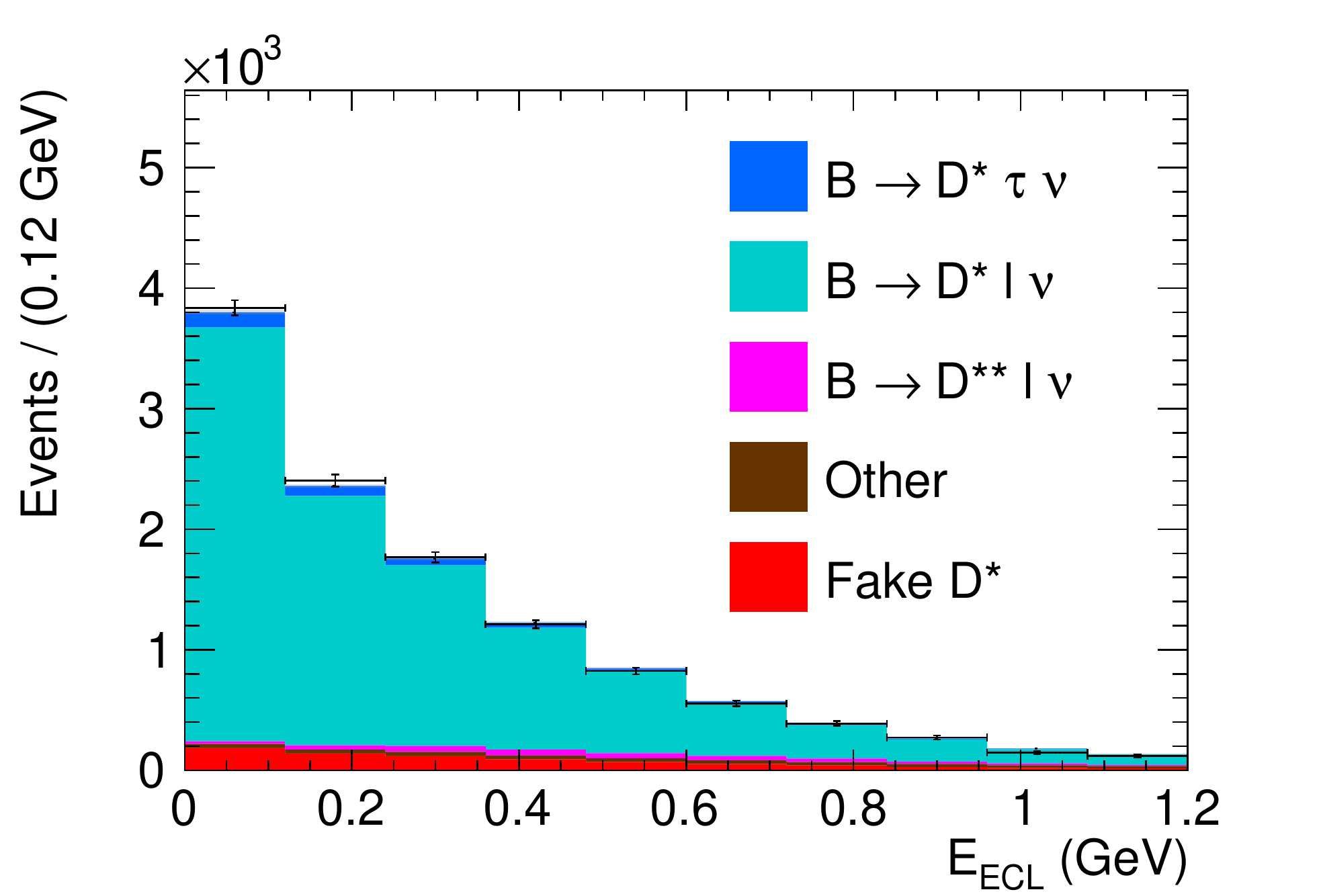}
	\includegraphics[width=\columnwidth]{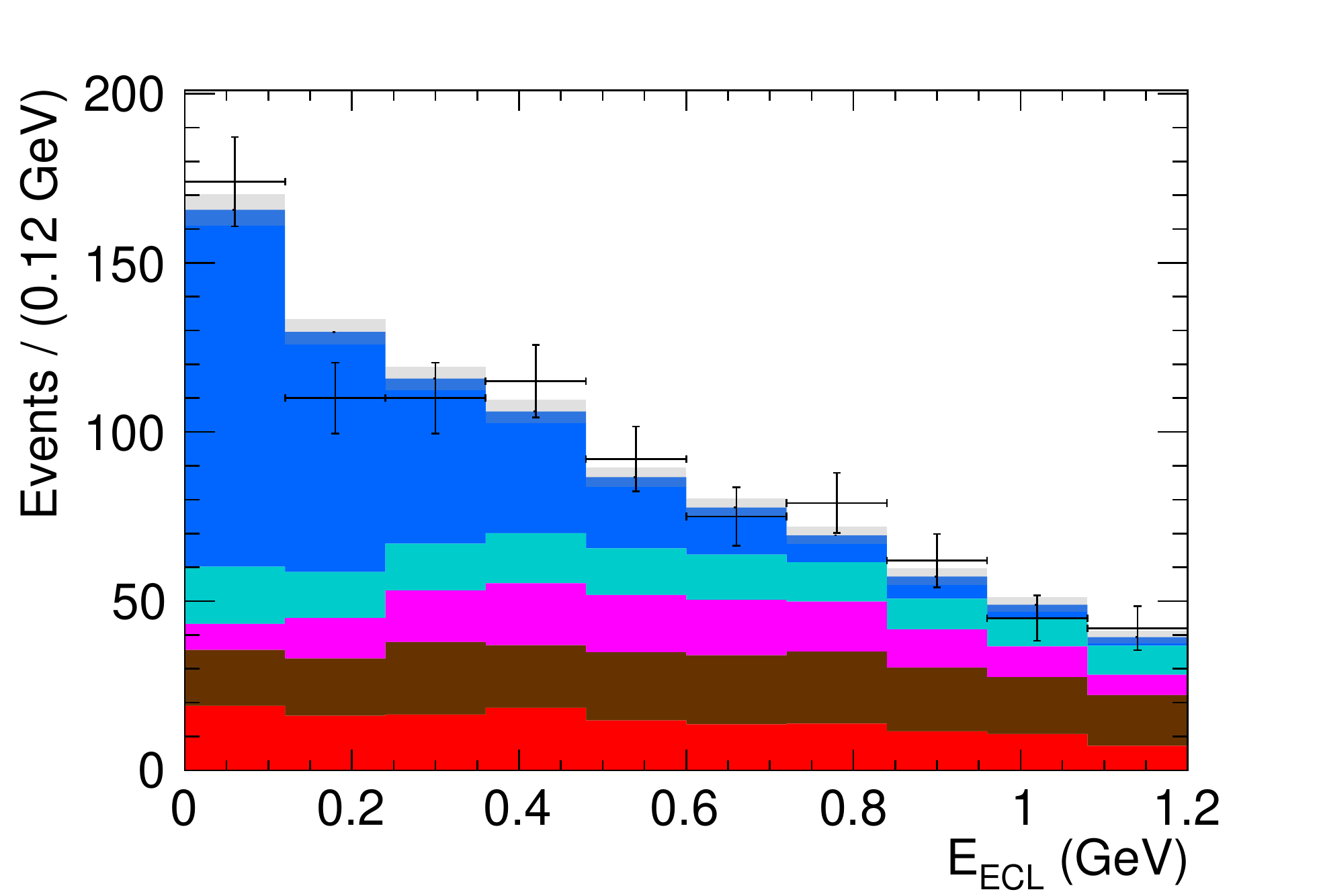}
	\includegraphics[width=\columnwidth]{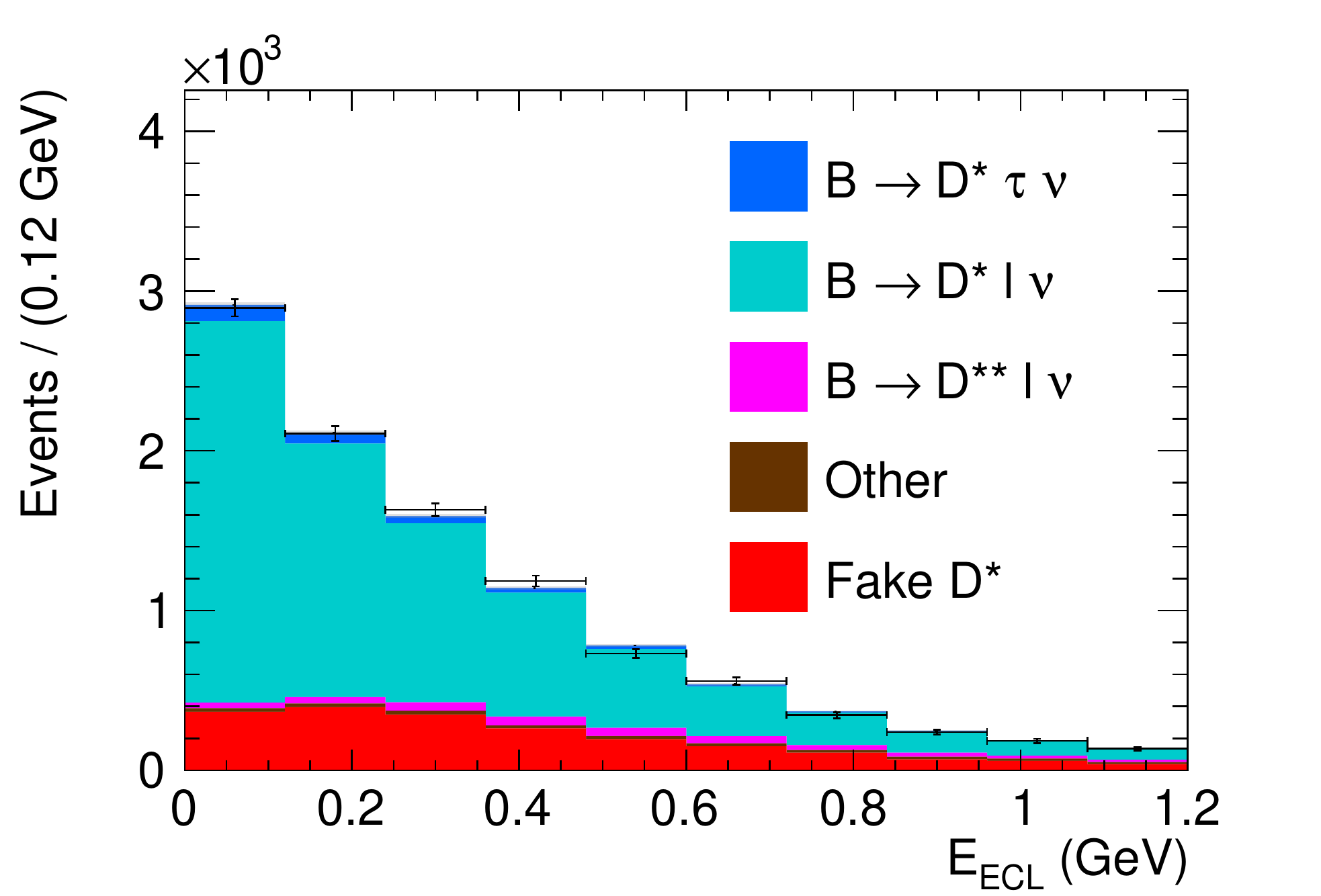}
	\includegraphics[width=\columnwidth]{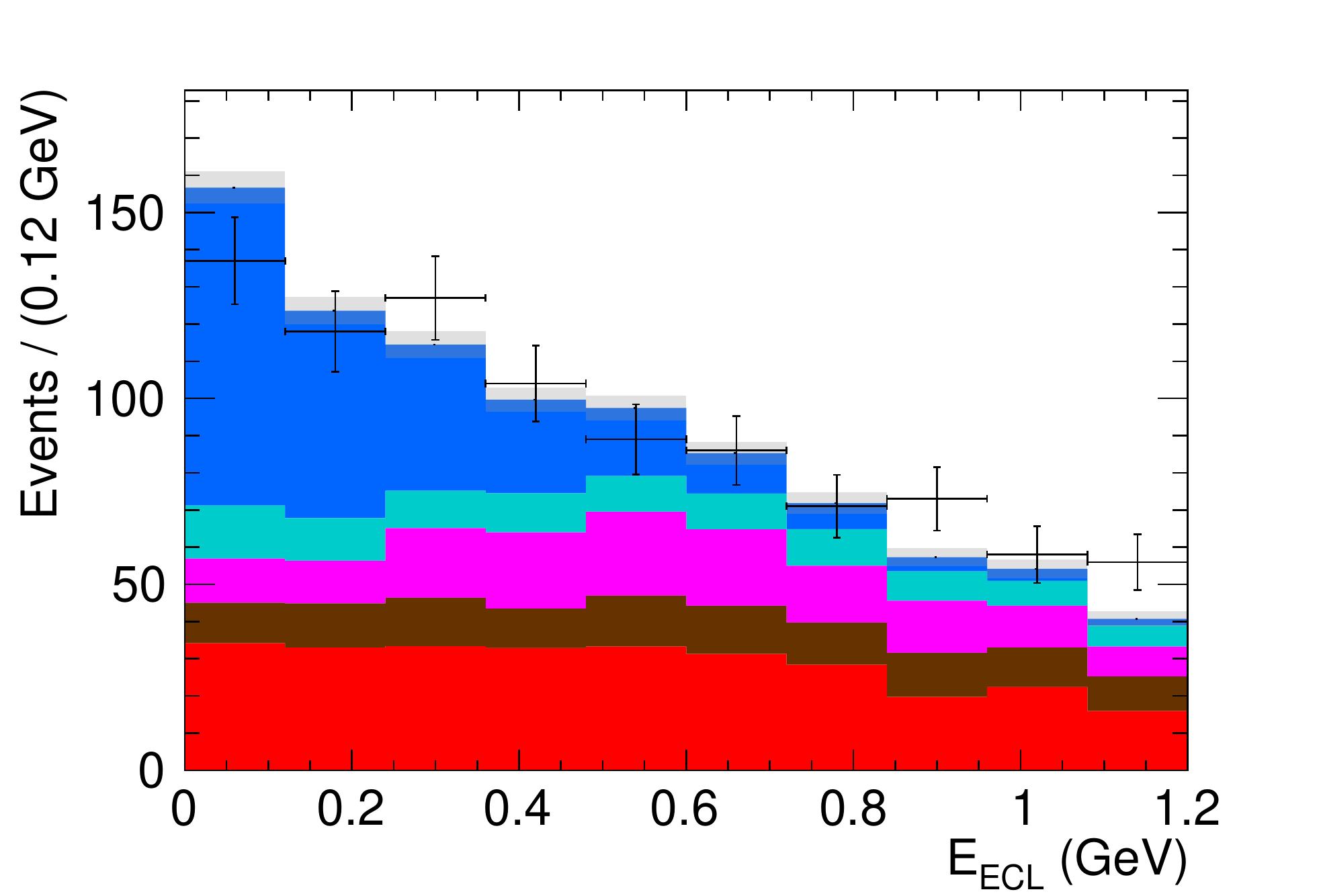}
\caption{\eecl\ fit projections and data points with statistical uncertainties in the $D^{\ast +}\ell^-$ (top) and $D^{\ast 0}\ell^-$ (bottom) samples, are shown for the full classifier region (left) and the signal region defined by the selection \texttt{class} $>0.9$ (right).}
\label{fig:results_Dstmodes}
\end{figure*}

\begin{figure*}[tbp]
	\includegraphics[width=\columnwidth]{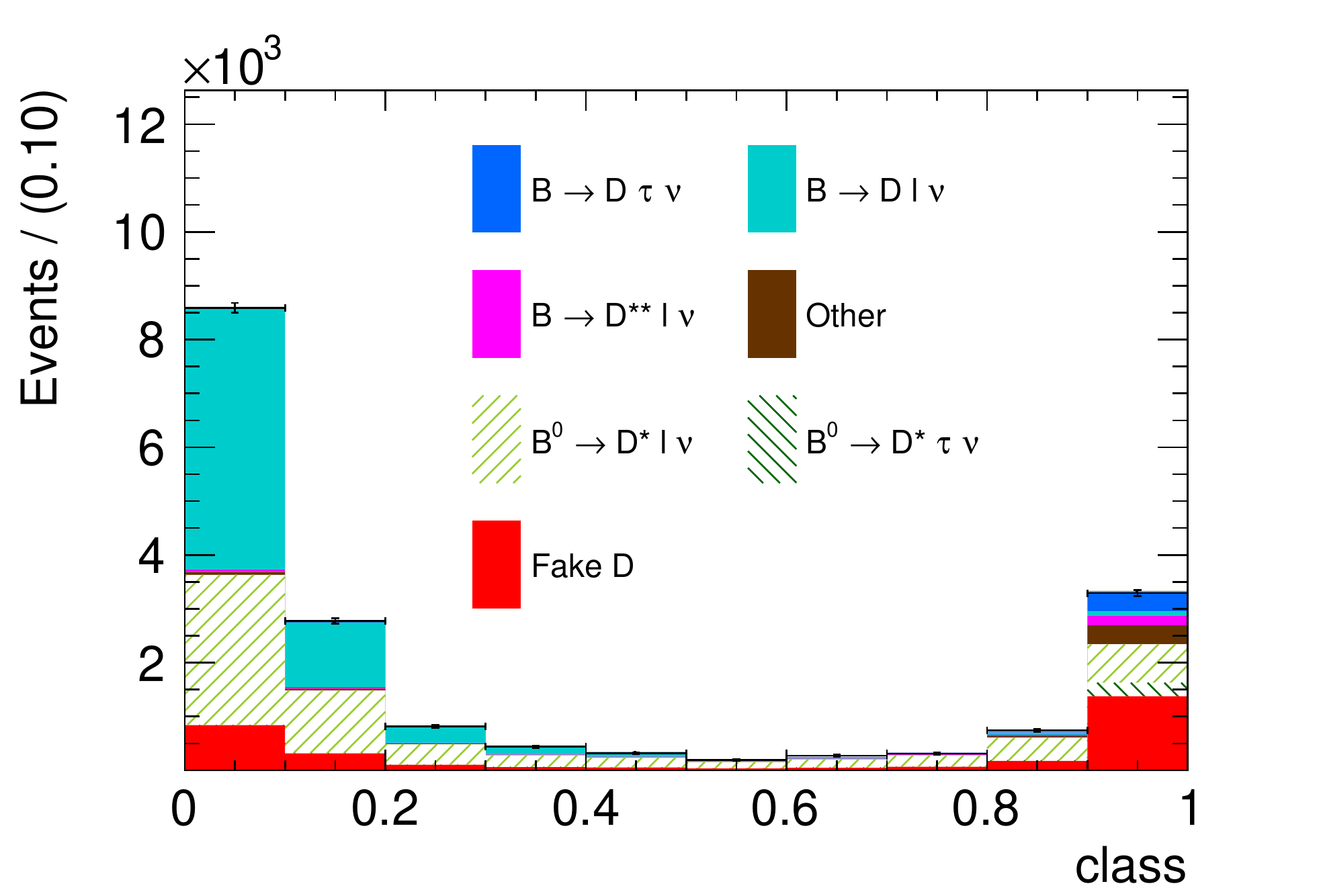}
	\includegraphics[width=\columnwidth]{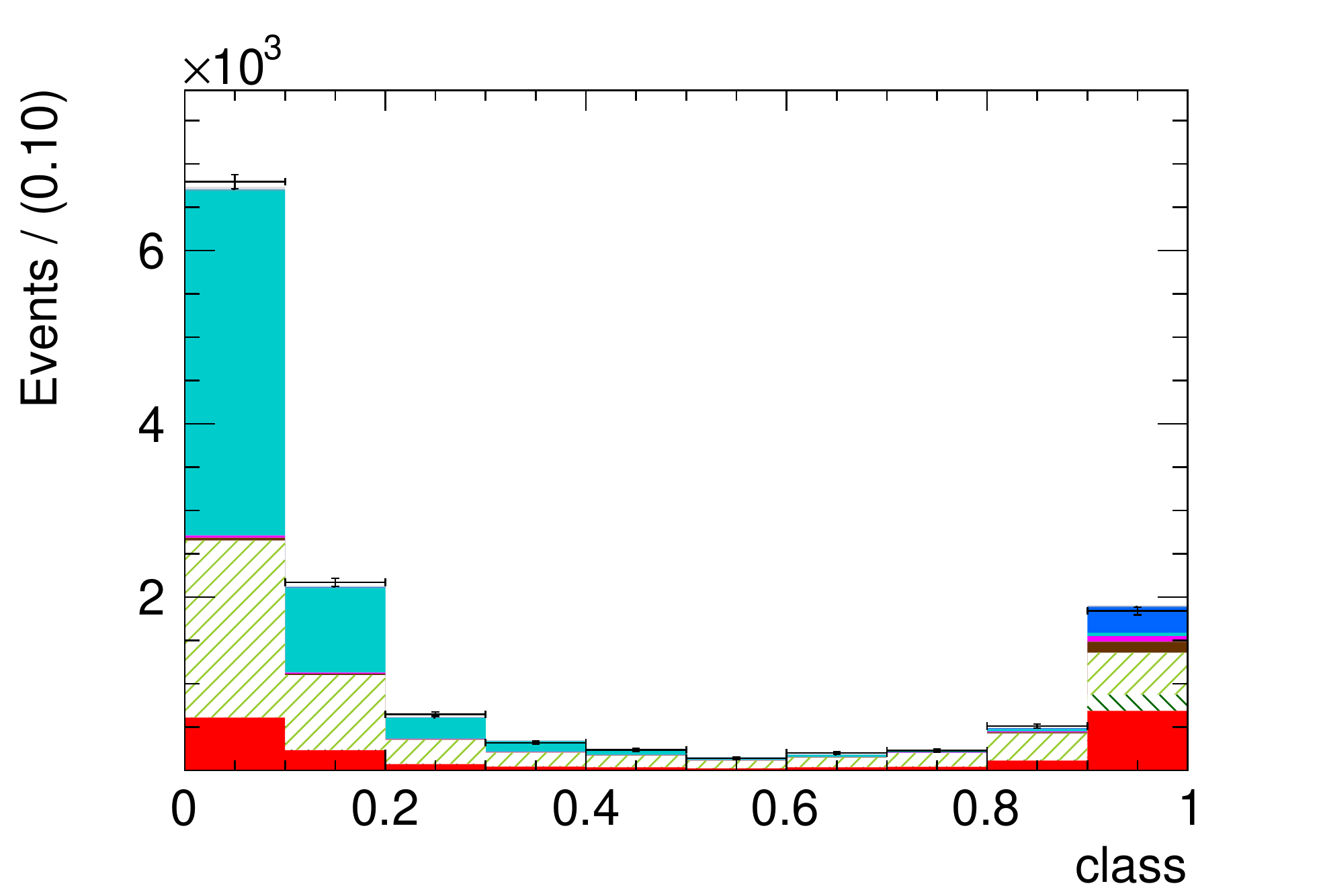}
	\includegraphics[width=\columnwidth]{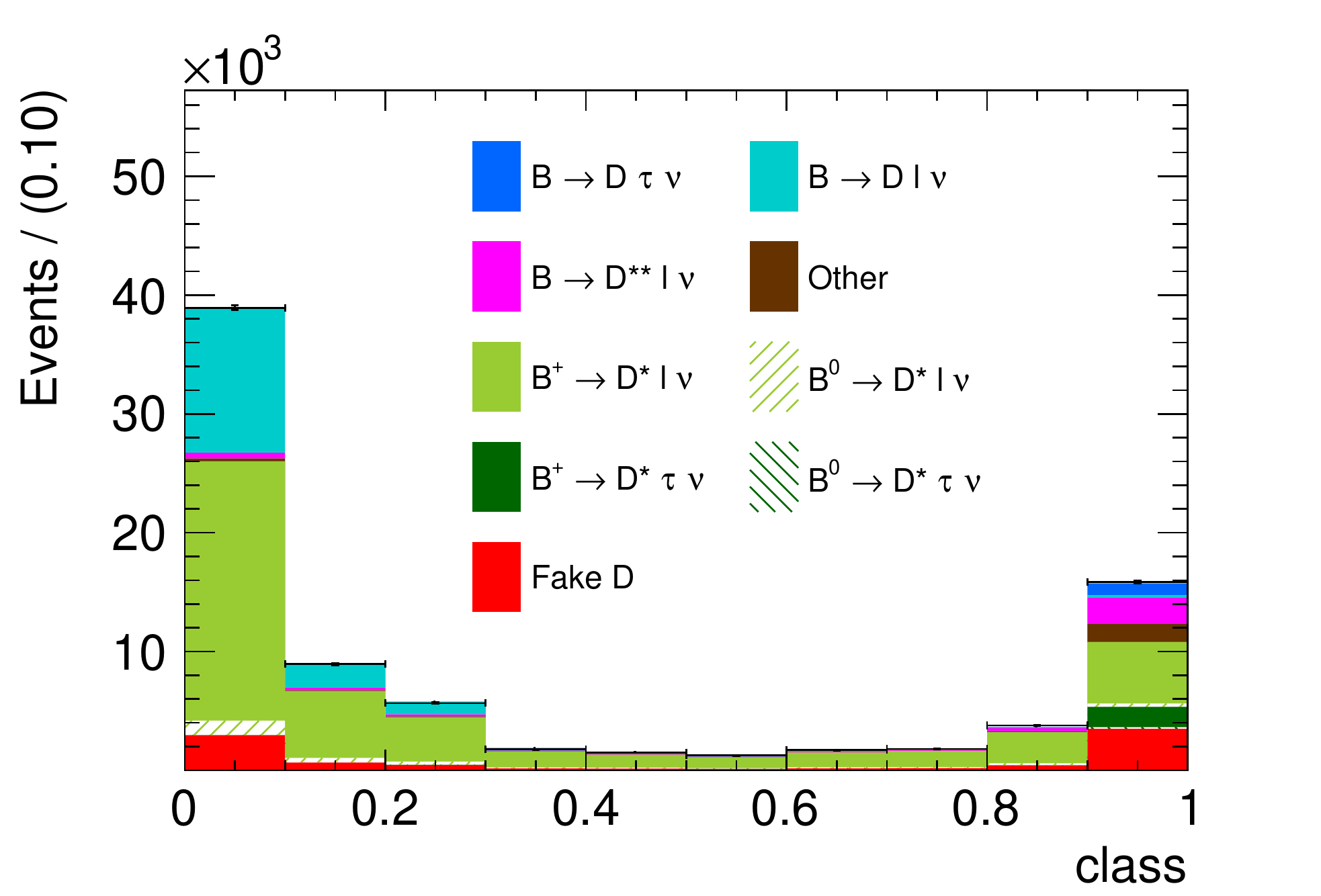}
	\includegraphics[width=\columnwidth]{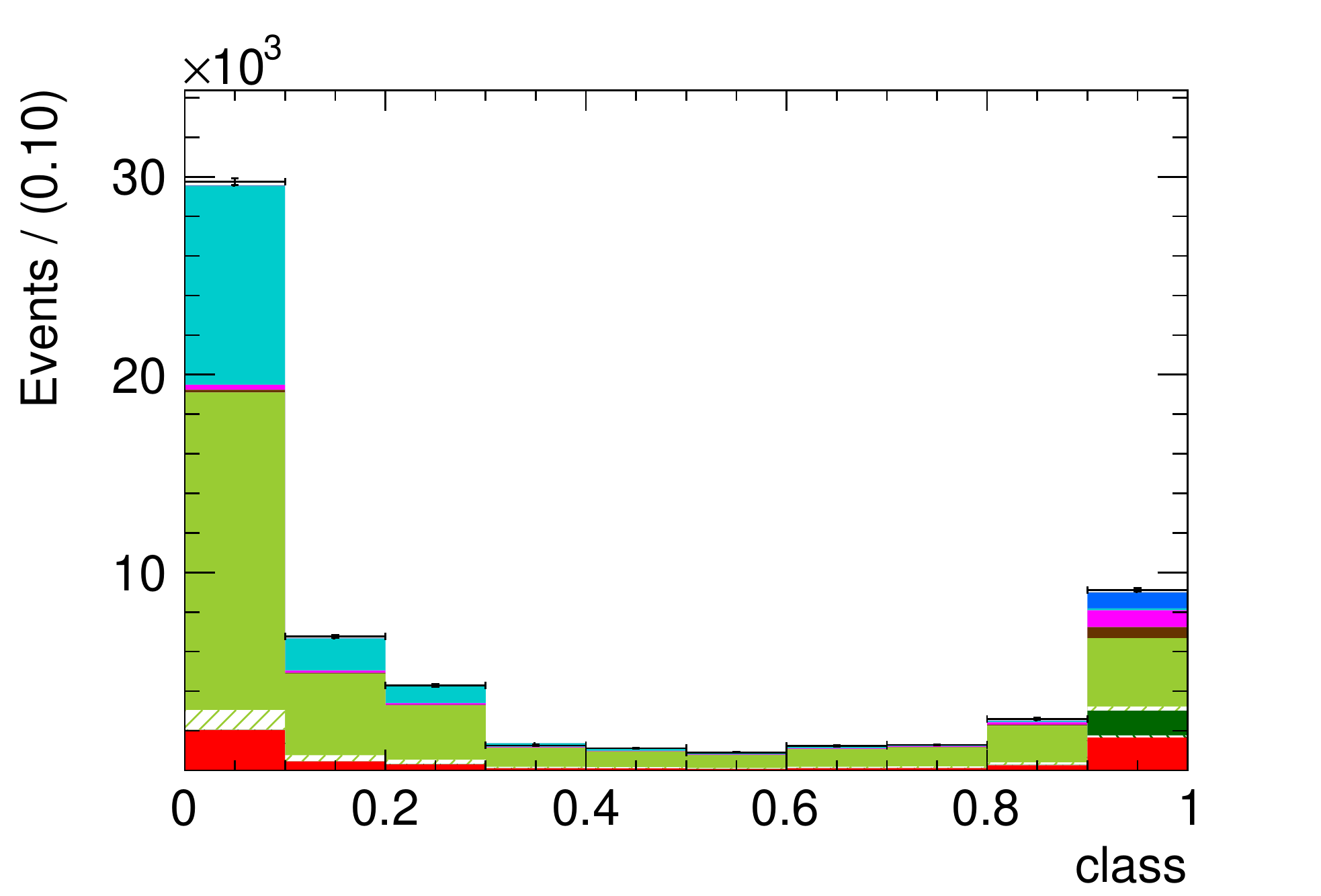}
\caption{Classifier fit projections and data points with statistical uncertainties in the $D^+\ell^-$ (top) and $D^0\ell^-$ (bottom) samples, are shown for the full \eecl\ region (left) and the signal region defined by the selection $\eecl < 0.48$ GeV (right).}
\label{fig:results_Dmodes_class}
\end{figure*}

\begin{figure*}[tbp]
	\includegraphics[width=\columnwidth]{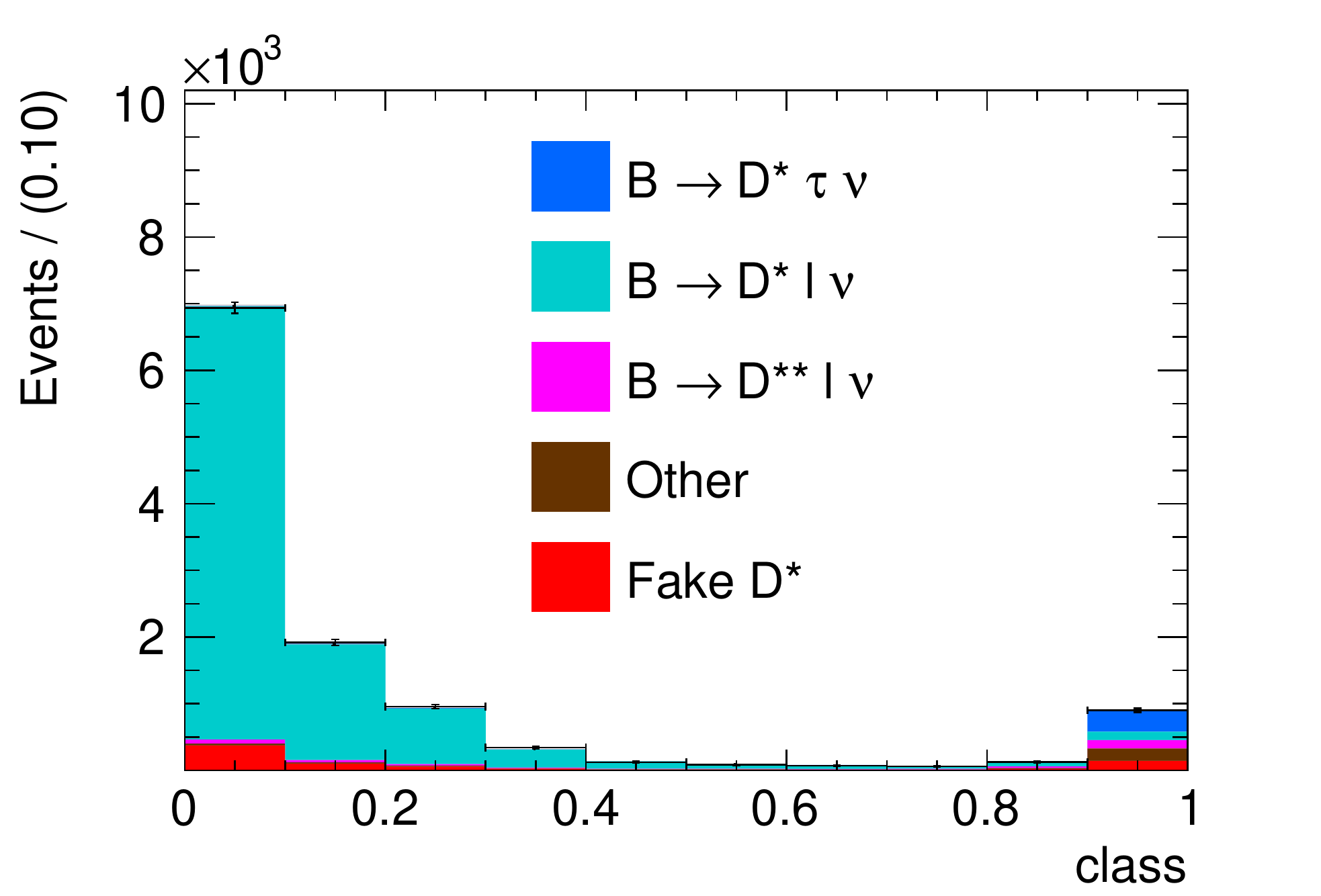}
	\includegraphics[width=\columnwidth]{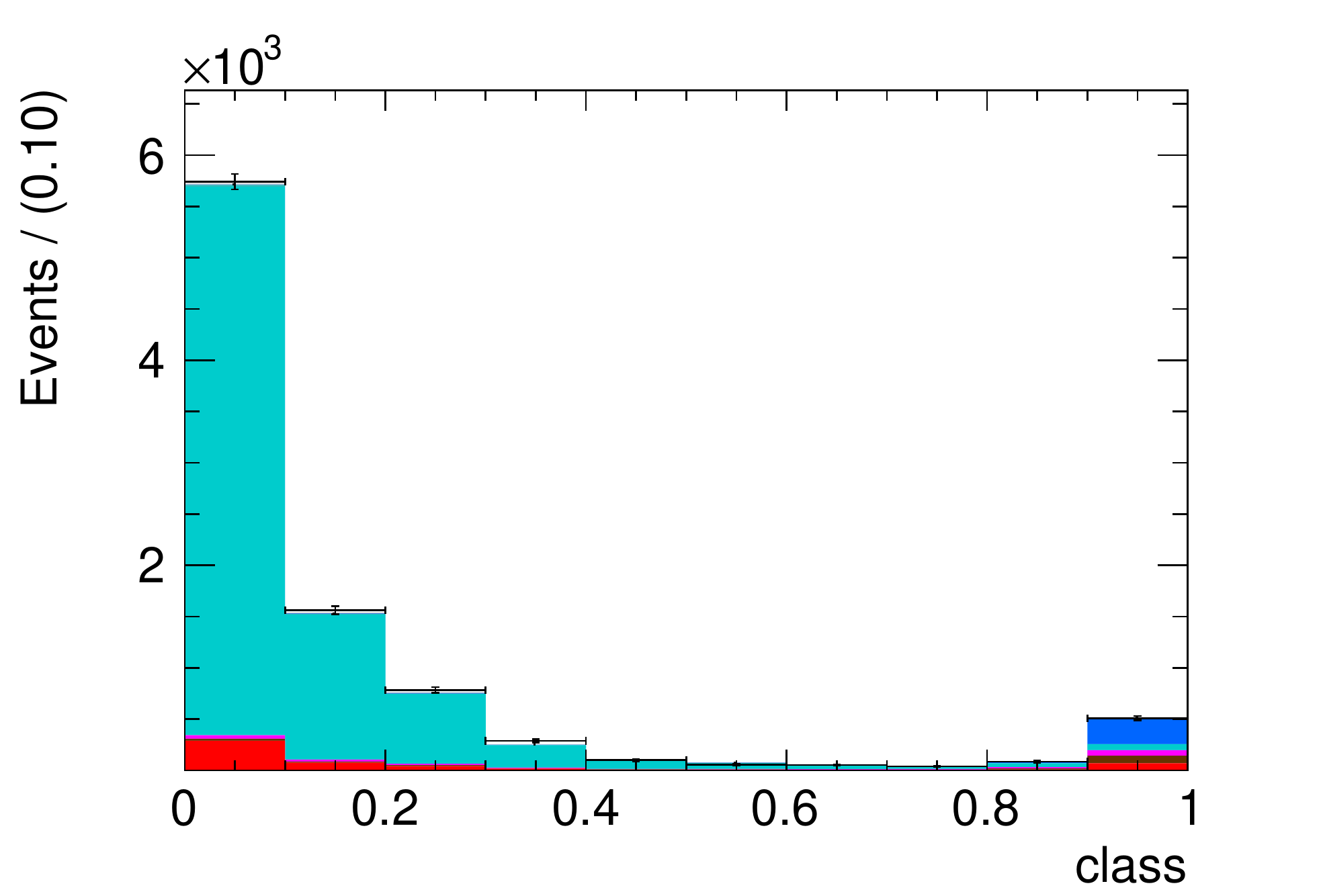}
	\includegraphics[width=\columnwidth]{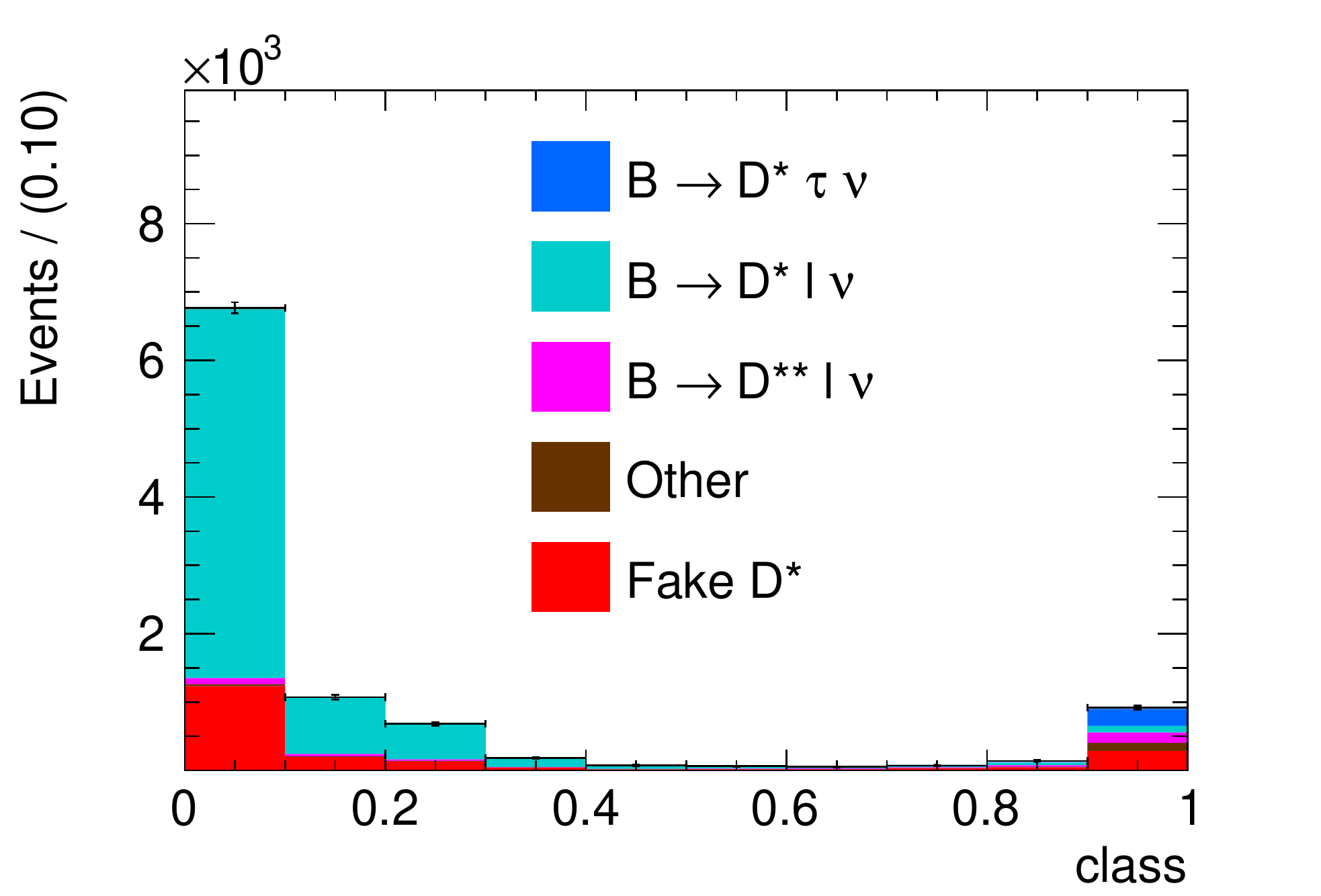}
	\includegraphics[width=\columnwidth]{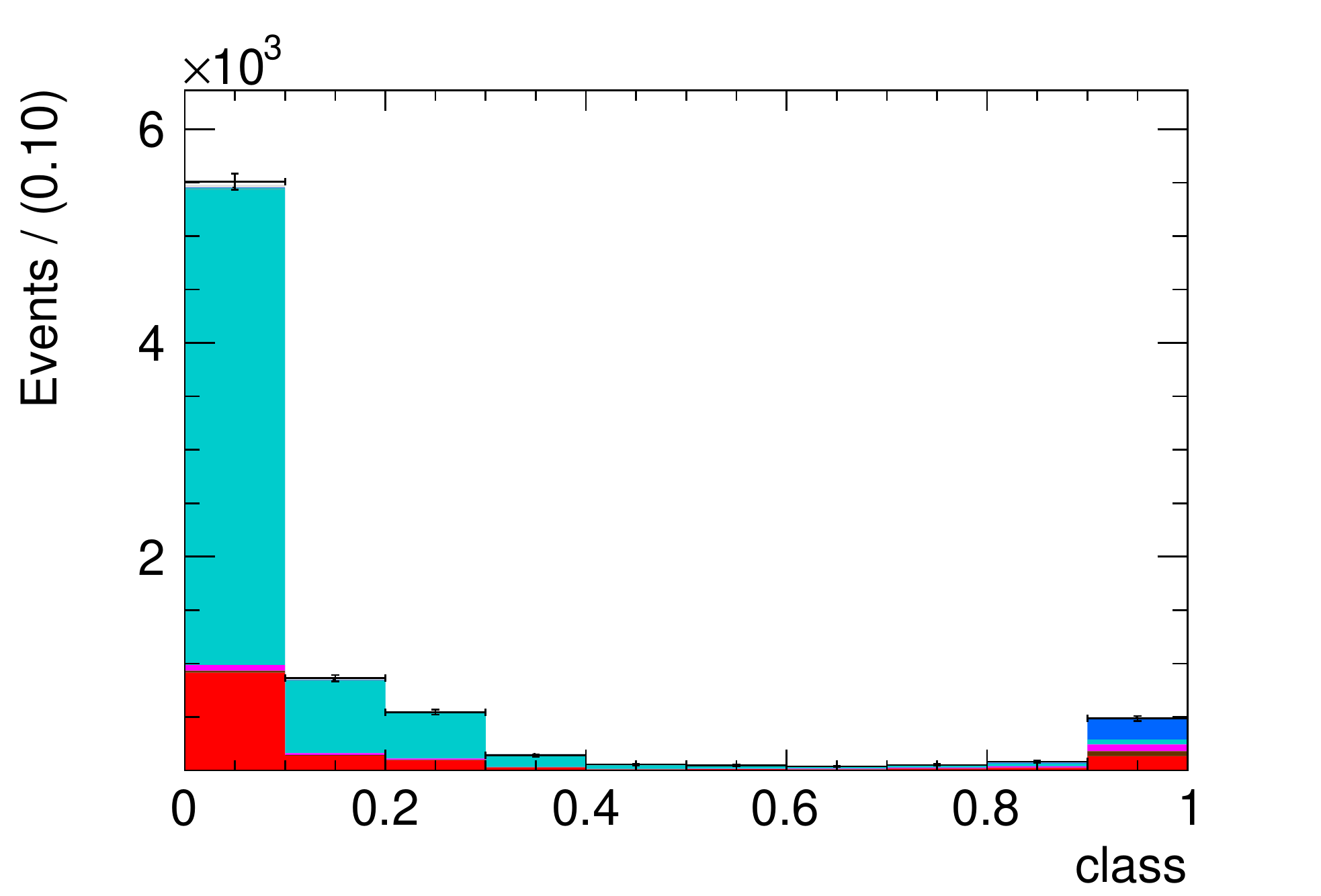}
\caption{Classifier fit projections and data points with statistical uncertainties in the $D^{\ast +}\ell^-$ (top) and $D^{\ast 0}\ell^-$ (bottom) samples, are shown for the full \eecl\ region (left) and the signal region defined by the selection $\eecl < 0.48$ GeV (right). }
\label{fig:results_Dstmodes_class}
\end{figure*}

\begin{figure*}[tbp]
	\includegraphics[width=\columnwidth]{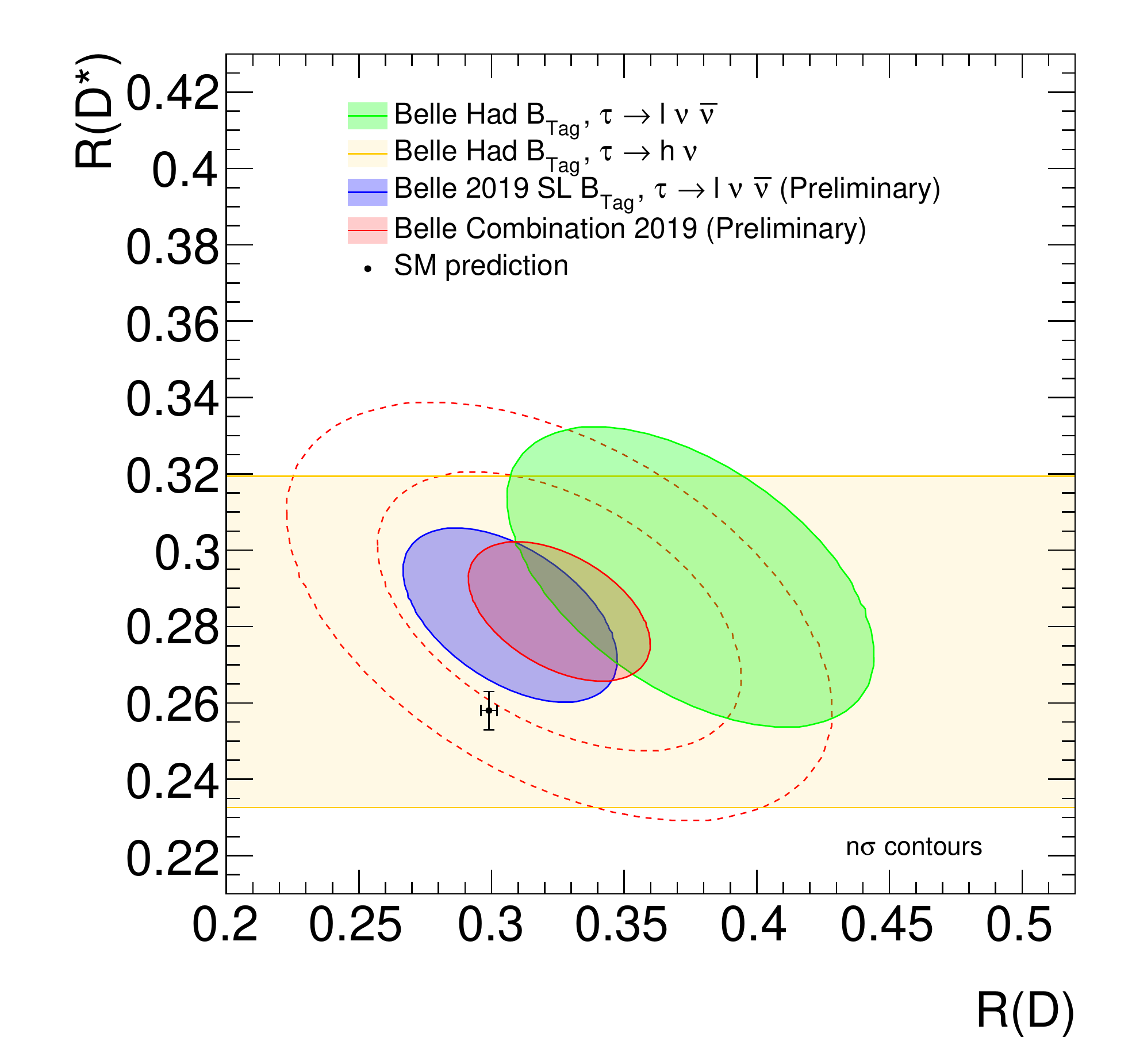}
    \caption{Fit results are shown on a $R(D)$ vs. $R(D^*)$ plane together with the most up-to-date Belle results on \RD\ and \RDSt\ performed with an hadronic tag~\cite{huschle2015, Hirose:2016wfn}. The latter results are combined with this measurement to provide the preliminary Belle combination, also shown in the plot. The yellow band and the blue and green ellipses are 1$\sigma$ contours, while red ellipses provide averaged results up to 3$\sigma$ contours. The SM expectations have values $\RD{}_{\text{SM}}  = 0.299 \pm 0.003$ and $\RDSt{}_{\text{SM}} = 0.258 \pm 0.005$~\cite{HFLAV}.}
    \label{fig:2Dplane}
\end{figure*}

Our results are:
\begin{align}
	\RD 	&= 0.307 \pm 0.037 \pm 0.016 \\
	\RDSt   &= 0.283 \pm 0.018 \pm 0.014,
\end{align}
where the first uncertainties are statistical, and the second are systematic. The same ordering of uncertainties holds for all following results. The statistical correlation between the quoted \RD\ and \RDSt\ values is $-0.53$, while the systematic correlation is $-0.52$. The dataset used in this measurement includes the one used for the previous $\R(D^{*+})$ result from Belle~\cite{Sato:2016svk}, which is consistent with this measurement. Being statistically correlated, the earlier measurement should not be averaged with this one, which combines $\R(D^{*+})$ and $\R(D^{*0})$. A breakdown of electron and muon channels yields $\R(D) = 0.281 \pm 0.042 \pm 0.017,\ \R(D^*)= 0.304 \pm 0.022 \pm 0.016$ for the first case, and $\R(D) = 0.373 \pm 0.068 \pm 0.030,\  \R(D^*) = 0.245 \pm 0.035 \pm 0.020$ for the second case. All fitted yields are listed in Table~\ref{tab:resultsYields}. The \eecl\ and \verb|class| projections of the fit are shown in Figures~\ref{fig:results_Dmodes},~\ref{fig:results_Dstmodes},~\ref{fig:results_Dmodes_class},~\ref{fig:results_Dstmodes_class}.
The 2D combination of the \RD\ and \RDSt\ results of this analysis, together with the most recent Belle results on \RD\ and \RDSt\ (\cite{huschle2015, Hirose:2016wfn}) obtained using a hadronic tag, are shown in Figure~\ref{fig:2Dplane}. The latter results are combined with this measurement to provide the preliminary Belle combination, which yields $\R(D) = 0.326 \pm 0.034,\ \R(D^*)= 0.284 \pm 0.018$ with a correlation equal to $-0.47$ between the \RD\ and \RDSt\ values. The preliminary Belle combination is also shown in Figure~\ref{fig:2Dplane}, with contours up to $3\sigma$.
\begin{table}[htb]
	\caption{Fit results for the yields of all components.} 
	\label{tab:resultsYields}
	\begin{ruledtabular}
    \centering
        \begin{tabular}{llc} 
            Channel             & Component                  & Yield \\ 
			\hline 
			$D^{+} \ell^{-}  $  & $B \to D \tau \nu$         &  $307 \pm 65$ \\ 
                    			& $B \to D \ell \nu$         &  $6800 \pm 179$ \\ 
                    			& $B^{0} \to D^{*} \ell \nu$ &  $6370 \pm 225$ \\ 
                    			& $B^{0} \to D^{*} \tau \nu$ &  $269 \pm 24$ \\ 
                    			& $B \to D^{**} \ell \nu $   &  $413 \pm 110$ \\ 
                    			& Fake $D$                   &  $3072 \pm 129$ (Fixed) \\ 
                    			& Other                      &  $506 \pm 23$ (Fixed) \\ 
            \hline 
			$D^{0} \ell^{-}$    & $B \to D \tau \nu$         &  $1471 \pm 193$ \\ 
                    			& $B \to D \ell \nu$         &  $16096 \pm 436$ \\ 
                    			& $B^{+} \to D^{*} \ell \nu$ &  $45042 \pm 563$ \\ 
                    			& $B^{0} \to D^{*} \ell \nu$ &  $2302 \pm 531$ \\ 
                    			& $B^{+} \to D^{*} \tau \nu$ &  $1704 \pm 177$ \\ 
                    			& $B^{0} \to D^{*} \tau \nu$ &  $123 \pm 11$ \\ 
                    			& $ B \to D^{**} \ell \nu $  &  $3595 \pm 252$ \\ 
                    			& Fake $D$                   &  $8708 \pm 418$ (Fixed) \\ 
                    			& Other                      &  $2131 \pm 83$ (Fixed) \\ 
			\hline
			$D^{*+} \ell^{-}  $ & $B \to D^{*} \tau \nu$     &  $376 \pm 36$ \\  
                    			& $B \to D^{*} \ell \nu$     &  $9794 \pm 109$ \\ 
                    			& $ B \to D^{**} \ell \nu $  &  $314 \pm 65$ \\ 
                    			& Fake $D^{*}$               &  $754 \pm 39$ (Fixed) \\ 
                    			& Other                      &  $287 \pm 13$ (Fixed) \\ 
			\hline
			$D^{*0} \ell^{-} $  & $B \to D^{*} \tau \nu$     &  $275 \pm 29$ \\ 
                    			& $B \to D^{*} \ell \nu$     &  $7148 \pm 100$ \\ 
                    			& $ B \to D^{**} \ell \nu $  &  $406 \pm 64$ \\ 
                    			& Fake $D^{*}$               &  $1993 \pm 122$ (Fixed) \\ 
                    			& Other                      &  $187 \pm 7$ (Fixed) \\ 
		\end{tabular} 
	\end{ruledtabular}
\end{table}

\section{CONCLUSION}

In summary, we have measured the ratios $
\RDall = {\cal B}(\bar{B} \to D^{(*)} \tau^- \bar{\nu}_{\tau})/{\cal B}(\bar{B} \to D^{(*)} \ell^- \bar{\nu}_{\ell})$, where $\ell$ denotes an electron or a muon, based on a semileptonic tagging method using a data sample containing $772 \times 10^6 B\bar{B}$ events collected with the Belle detector.
The results are
\begin{align}
	\RD 	&= 0.307 \pm 0.037 \pm 0.016 \\
	\RDSt   &= 0.283 \pm 0.018 \pm 0.014,
\end{align}
which are in agreement with the SM predictions within $0.2\sigma$ and $1.1\sigma$, respectively. The combined result agrees with the SM predictions within $1.2\sigma$. This work constitutes the most precise measurements of \RD\ and \RDSt\ performed to date. Furthermore, this is the first result for \RD\ based on a semileptonic tagging method.

\section{ACKNOWLEDGEMENTS}

We thank the KEKB group for excellent operation of the
accelerator; the KEK cryogenics group for efficient solenoid
operations; and the KEK computer group, the NII, and 
PNNL/EMSL for valuable computing and SINET5 network support.  
We acknowledge support from MEXT, JSPS and Nagoya's TLPRC (Japan);
ARC (Australia); FWF (Austria); NSFC and CCEPP (China); 
MSMT (Czechia); CZF, DFG, EXC153, and VS (Germany);
DST (India); INFN (Italy); 
MOE, MSIP, NRF, RSRI, FLRFAS project and GSDC of KISTI and KREONET/GLORIAD (Korea);
MNiSW and NCN (Poland); MSHE (Russia); ARRS (Slovenia);
IKERBASQUE (Spain); 
SNSF (Switzerland); MOE and MOST (Taiwan); and DOE and NSF (USA).

\bibliography{bibliography}

\end{document}

%% file: imports/authors.tex
\noaffiliation
\affiliation{University of the Basque Country UPV/EHU, 48080 Bilbao}
\affiliation{Beihang University, Beijing 100191}
\affiliation{University of Bonn, 53115 Bonn}
\affiliation{Brookhaven National Laboratory, Upton, New York 11973}
\affiliation{Budker Institute of Nuclear Physics SB RAS, Novosibirsk 630090}
\affiliation{Faculty of Mathematics and Physics, Charles University, 121 16 Prague}
\affiliation{Chiba University, Chiba 263-8522}
\affiliation{Chonnam National University, Kwangju 660-701}
\affiliation{University of Cincinnati, Cincinnati, Ohio 45221}
\affiliation{Deutsches Elektronen--Synchrotron, 22607 Hamburg}
\affiliation{Duke University, Durham, North Carolina 27708}
\affiliation{University of Florida, Gainesville, Florida 32611}
\affiliation{Department of Physics, Fu Jen Catholic University, Taipei 24205}
\affiliation{Key Laboratory of Nuclear Physics and Ion-beam Application (MOE) and Institute of Modern Physics, Fudan University, Shanghai 200443}
\affiliation{Justus-Liebig-Universit\"at Gie\ss{}en, 35392 Gie\ss{}en}
\affiliation{Gifu University, Gifu 501-1193}
\affiliation{II. Physikalisches Institut, Georg-August-Universit\"at G\"ottingen, 37073 G\"ottingen}
\affiliation{SOKENDAI (The Graduate University for Advanced Studies), Hayama 240-0193}
\affiliation{Gyeongsang National University, Chinju 660-701}
\affiliation{Hanyang University, Seoul 133-791}
\affiliation{University of Hawaii, Honolulu, Hawaii 96822}
\affiliation{High Energy Accelerator Research Organization (KEK), Tsukuba 305-0801}
\affiliation{J-PARC Branch, KEK Theory Center, High Energy Accelerator Research Organization (KEK), Tsukuba 305-0801}
\affiliation{Forschungszentrum J\"{u}lich, 52425 J\"{u}lich}
\affiliation{Hiroshima Institute of Technology, Hiroshima 731-5193}
\affiliation{IKERBASQUE, Basque Foundation for Science, 48013 Bilbao}
\affiliation{University of Illinois at Urbana-Champaign, Urbana, Illinois 61801}
\affiliation{Indian Institute of Science Education and Research Mohali, SAS Nagar, 140306}
\affiliation{Indian Institute of Technology Bhubaneswar, Satya Nagar 751007}
\affiliation{Indian Institute of Technology Guwahati, Assam 781039}
\affiliation{Indian Institute of Technology Hyderabad, Telangana 502285}
\affiliation{Indian Institute of Technology Madras, Chennai 600036}
\affiliation{Indiana University, Bloomington, Indiana 47408}
\affiliation{Institute of High Energy Physics, Chinese Academy of Sciences, Beijing 100049}
\affiliation{Institute of High Energy Physics, Vienna 1050}
\affiliation{Institute for High Energy Physics, Protvino 142281}
\affiliation{Institute of Mathematical Sciences, Chennai 600113}
\affiliation{INFN - Sezione di Napoli, 80126 Napoli}
\affiliation{INFN - Sezione di Torino, 10125 Torino}
\affiliation{Advanced Science Research Center, Japan Atomic Energy Agency, Naka 319-1195}
\affiliation{J. Stefan Institute, 1000 Ljubljana}
\affiliation{Kanagawa University, Yokohama 221-8686}
\affiliation{Institut f\"ur Experimentelle Teilchenphysik, Karlsruher Institut f\"ur Technologie, 76131 Karlsruhe}
\affiliation{Kavli Institute for the Physics and Mathematics of the Universe (WPI), University of Tokyo, Kashiwa 277-8583}
\affiliation{Kennesaw State University, Kennesaw, Georgia 30144}
\affiliation{King Abdulaziz City for Science and Technology, Riyadh 11442}
\affiliation{Department of Physics, Faculty of Science, King Abdulaziz University, Jeddah 21589}
\affiliation{Kitasato University, Sagamihara 252-0373}
\affiliation{Korea Institute of Science and Technology Information, Daejeon 305-806}
\affiliation{Korea University, Seoul 136-713}
\affiliation{Kyoto University, Kyoto 606-8502}
\affiliation{Kyungpook National University, Daegu 702-701}
\affiliation{LAL, Univ. Paris-Sud, CNRS/IN2P3, Universit\'{e} Paris-Saclay, Orsay}
\affiliation{\'Ecole Polytechnique F\'ed\'erale de Lausanne (EPFL), Lausanne 1015}
\affiliation{P.N. Lebedev Physical Institute of the Russian Academy of Sciences, Moscow 119991}
\affiliation{Liaoning Normal University, Dalian 116029}
\affiliation{Faculty of Mathematics and Physics, University of Ljubljana, 1000 Ljubljana}
\affiliation{Ludwig Maximilians University, 80539 Munich}
\affiliation{Luther College, Decorah, Iowa 52101}
\affiliation{Malaviya National Institute of Technology Jaipur, Jaipur 302017}
\affiliation{University of Malaya, 50603 Kuala Lumpur}
\affiliation{University of Maribor, 2000 Maribor}
\affiliation{Max-Planck-Institut f\"ur Physik, 80805 M\"unchen}
\affiliation{School of Physics, University of Melbourne, Victoria 3010}
\affiliation{University of Mississippi, University, Mississippi 38677}
\affiliation{University of Miyazaki, Miyazaki 889-2192}
\affiliation{Moscow Physical Engineering Institute, Moscow 115409}
\affiliation{Moscow Institute of Physics and Technology, Moscow Region 141700}
\affiliation{Graduate School of Science, Nagoya University, Nagoya 464-8602}
\affiliation{Kobayashi-Maskawa Institute, Nagoya University, Nagoya 464-8602}
\affiliation{Universit\`{a} di Napoli Federico II, 80055 Napoli}
\affiliation{Nara University of Education, Nara 630-8528}
\affiliation{Nara Women's University, Nara 630-8506}
\affiliation{National Central University, Chung-li 32054}
\affiliation{National United University, Miao Li 36003}
\affiliation{Department of Physics, National Taiwan University, Taipei 10617}
\affiliation{H. Niewodniczanski Institute of Nuclear Physics, Krakow 31-342}
\affiliation{Nippon Dental University, Niigata 951-8580}
\affiliation{Niigata University, Niigata 950-2181}
\affiliation{University of Nova Gorica, 5000 Nova Gorica}
\affiliation{Novosibirsk State University, Novosibirsk 630090}
\affiliation{Osaka City University, Osaka 558-8585}
\affiliation{Osaka University, Osaka 565-0871}
\affiliation{Pacific Northwest National Laboratory, Richland, Washington 99352}
\affiliation{Panjab University, Chandigarh 160014}
\affiliation{Peking University, Beijing 100871}
\affiliation{University of Pittsburgh, Pittsburgh, Pennsylvania 15260}
\affiliation{Punjab Agricultural University, Ludhiana 141004}
\affiliation{Research Center for Electron Photon Science, Tohoku University, Sendai 980-8578}
\affiliation{Research Center for Nuclear Physics, Osaka University, Osaka 567-0047}
\affiliation{Theoretical Research Division, Nishina Center, RIKEN, Saitama 351-0198}
\affiliation{RIKEN BNL Research Center, Upton, New York 11973}
\affiliation{Saga University, Saga 840-8502}
\affiliation{University of Science and Technology of China, Hefei 230026}
\affiliation{Seoul National University, Seoul 151-742}
\affiliation{Shinshu University, Nagano 390-8621}
\affiliation{Showa Pharmaceutical University, Tokyo 194-8543}
\affiliation{Soongsil University, Seoul 156-743}
\affiliation{University of South Carolina, Columbia, South Carolina 29208}
\affiliation{Stefan Meyer Institute for Subatomic Physics, Vienna 1090}
\affiliation{Sungkyunkwan University, Suwon 440-746}
\affiliation{School of Physics, University of Sydney, New South Wales 2006}
\affiliation{Department of Physics, Faculty of Science, University of Tabuk, Tabuk 71451}
\affiliation{Tata Institute of Fundamental Research, Mumbai 400005}
\affiliation{Excellence Cluster Universe, Technische Universit\"at M\"unchen, 85748 Garching}
\affiliation{Department of Physics, Technische Universit\"at M\"unchen, 85748 Garching}
\affiliation{Toho University, Funabashi 274-8510}
\affiliation{Tohoku Gakuin University, Tagajo 985-8537}
\affiliation{Department of Physics, Tohoku University, Sendai 980-8578}
\affiliation{Earthquake Research Institute, University of Tokyo, Tokyo 113-0032}
\affiliation{Department of Physics, University of Tokyo, Tokyo 113-0033}
\affiliation{Tokyo Institute of Technology, Tokyo 152-8550}
\affiliation{Tokyo Metropolitan University, Tokyo 192-0397}
\affiliation{Tokyo University of Agriculture and Technology, Tokyo 184-8588}
\affiliation{Utkal University, Bhubaneswar 751004}
\affiliation{Virginia Polytechnic Institute and State University, Blacksburg, Virginia 24061}
\affiliation{Wayne State University, Detroit, Michigan 48202}
\affiliation{Yamagata University, Yamagata 990-8560}
\affiliation{Yonsei University, Seoul 120-749}
  \author{A.~Abdesselam}\affiliation{Department of Physics, Faculty of Science, University of Tabuk, Tabuk 71451} 
  \author{I.~Adachi}\affiliation{High Energy Accelerator Research Organization (KEK), Tsukuba 305-0801}\affiliation{SOKENDAI (The Graduate University for Advanced Studies), Hayama 240-0193} 
  \author{K.~Adamczyk}\affiliation{H. Niewodniczanski Institute of Nuclear Physics, Krakow 31-342} 
  \author{J.~K.~Ahn}\affiliation{Korea University, Seoul 136-713} 
  \author{H.~Aihara}\affiliation{Department of Physics, University of Tokyo, Tokyo 113-0033} 
  \author{S.~Al~Said}\affiliation{Department of Physics, Faculty of Science, University of Tabuk, Tabuk 71451}\affiliation{Department of Physics, Faculty of Science, King Abdulaziz University, Jeddah 21589} 
  \author{K.~Arinstein}\affiliation{Budker Institute of Nuclear Physics SB RAS, Novosibirsk 630090}\affiliation{Novosibirsk State University, Novosibirsk 630090} 
  \author{Y.~Arita}\affiliation{Graduate School of Science, Nagoya University, Nagoya 464-8602} 
  \author{D.~M.~Asner}\affiliation{Brookhaven National Laboratory, Upton, New York 11973} 
  \author{H.~Atmacan}\affiliation{University of South Carolina, Columbia, South Carolina 29208} 
% ...
\collaboration{The Belle Collaboration}

%% file: main_paper.bbl
\begin{thebibliography}{28}
\bibitem [{not()}]{note}%
  \BibitemOpen
  \href@noop {} {}\bibinfo {note} {{{Throughout this paper, the inclusion of
  the charge-conjugate decay mode is implied}}}\BibitemShut {NoStop}%
\end{thebibliography}
